\documentstyle [12pt,epsf]{article}
\begin{document}
\newcommand{\vm}{\vspace{0.2cm}}
\newcommand{\vl}{\vspace{0.4cm}}

\title{Recent status of leptohadron hypothesis }

\author{Matti Pitk\"anen
\\
\small \em Torkkelinkatu 21 B39 , 00530, Helsinki, Finland\\
}

\date{16. October 1995}

\maketitle

\newpage

\tableofcontents
\begin{center}
{\bf Abstract}
\end{center}

\vm

TGD predicts the existence of color octet excitations of ordinary leptons
forming 'leptohadrons' as their color singlet bound states. There is some
evidence on the existence of leptohadrons: the production of anomalous $e^+e^-$
pairs in heavy ion collisions, Karmen anomaly, the anomalously high decay rate
of ortopositronium (Op) and anomalous production of low energy  $e^+e^-$ pairs
in hadronic collisions. PCAC  makes it possible to predict the couplings of
leptopion to leptons.  The new contribution to Op decay rate is of correct
order of magnitude and the anomaly allows to determine the precise value of the
parameter $f_{\pi_L}$. Sigma model realization of  PCAC makes it possible to
construct a model for the production of leptohadrons in the electromagnetic
fields of   the colliding nuclei. $\pi_L$ develops vacuum expectation value
proportional to the 'instanton density' $E\cdot B$ and it is possible to relate
leptohadron production  rates  to the Fourier transform of $E\cdot B$.
Anomalous $e^+e^-$ pairs must originate from $\sigma_L \pi_L$ pairs  via
$\sigma_L\rightarrow e^+e^-$ decay. The peculiar production characteristics  of
leptomesons  are reproduced, the order of magnitude for the production  cross
section is correct and various decay rates are within experimental bounds. A
resonance in  photon photon scattering at cm energy equal to leptopion mass is
predicted  and leptobaryon pair production in heavy ion collisions is in
principle possible. Leptopion  contribution to the $\nu-e$ and $\bar{\nu}-e$
scattering should dominate at low energies.

\newpage

\section{Introduction}

TGD suggest strongly  ('predicts' is perhaps too strong expression)
the existence of color excited leptons.
The mass calculations based on p-adic thermodynamics and p-adic conformal
invariance lead to a rather detailed picture about color excited leptons
\cite{padmasses}.\\
a) Color excited leptons are  color octets and leptohadrons are formed
as their color
singlet bound states.\\
b) The basic  mass scale for leptohadron physics
is completely fixed by the assumption that
leptohadrons
correspond to condensate level $M_{127}$ (for the p-adic formulation of
 topological
condensate concept see \cite{padTGD}).   Color
excited leptons can have  $k=127,113,107,...$ ($p\simeq 2^k$, $k $ prime)
 condensation levels as primary
condensation levels.
The mass spectrum of leptohadrons
is expected to have same general characteristics as hadronic mass spectrum
and a satisfactory description should be based on string tension concept.
The masses of ground state leptohadrons are calculable once
primary condensation levels for colored leptons and  the CKM matrix
describing the mixing of color excited lepton families is known.

\vm

The strongest counter arguments against color excited leptons are
 the following
 ones:
\\
a) The decay width of $Z^0$ boson allows only $N=3$ light particles
with neutrino quantum numbers. The introduction of new light elementary
 particles
makes the decay width of $Z^0$ untolerably large. A purely TGD:eish solution
of the problem was  proposed on \cite{padmasses} (5:th paper of series)
 and
relied heavily on
 the relationship between p-adic and real probabilities.\\
b) The introduction of new colored states  (also exotic quarks)  spoils the
asymptotic freedom
of QCD. The proposed solution of problem was based on the idea that there is
a different QCD associated with each  Mersenne prime and these QCD:s do not
 communicate with
each other.  Also  colored exotic bosons are predicted and these save the
asymptotic freedom for each QCD.

\vm

One might stop the reading after these counterarguments unless there
were definite experimental evidence supporting the leptohadron hypothesis.\\
a) The production of anomalous $e^+e^-$ pairs in heavy ion collisions
(energies just above the Coulomb barrier) suggests the existence of
pseudoscalar particles decaying to $e^+e^-$ pairs. In \cite{Lepto} these
states were identified as leptopions that is bound states of color octet
 excitations
of $e^+$ and $e^-$. The model for leptopion  production
was based on PCAC argument and led to an explanation for the peculiar
production
characteristics of leptopion.   In  \cite{Heavy} the model was developed
further by applying PCAC hypothesis. Unfortumately the calculations
contained some (rather stupid) errors
and led to too optimistic conclusions and partially erranous physical picture.
  For instance, the  predicted production
 cross
 section was found  to be
of correct order of magnitude: unfortunately this was due to an error in
 numerical
calculation. \\
b) The second puzzle, Karmen anomaly, is quite recent \cite{Karmen}.
It has been found that in charge pion decay the  distribution for
the number of neutrinos accompanying muon in  decay $\pi \rightarrow \mu +
\nu_{\mu}$ as a function of time seems to have a small shoulder at
$t_0\sim ms$. A possible explanation is the decay of charged pion
to muon plus some new weakly interacting particle with mass of order
 $30 \ MeV$
\cite{Karmen1}: the production and decay of this particle would proceed
via mixing with muon neutrino.  TGD suggests the identification of this
state as leptobaryon of type $L_B=f_{abc}L_8^aL_8^b\bar{L}_8^c$
having electroweak quantum numbers of neutrino. The mass  of the exotic
 neutrino
is  indeed of  correct order of magnitude (given by the muon mass scale).\\
c) The third puzzle is the  anomalously high decay rate of ortopositronium.
\cite{orto}.  $e^+e^-$ annihilation to
virtual photon followed by the decay to real photon plus virtual
leptopion followed by the decay of the  virtual leptopion to real photon
pair,
 $\pi_L\gamma\gamma$ coupling being
determined by  axial anomaly, provides a possible explanation of the
puzzle. \\
d) There is also an anomalously large production of low energy
$e^+e^-$ pairs \cite{anopair,Barshay} in hadronic collisions, which might be
basically due to the production of leptohadrons via the decay of virtual
photons to colored leptons.

 \vm

In this chapter a revised form   of leptohadron hypothesis is described.\\
a) Sigma model realization of PCAC hypothesis allows to determine the decay
 widths of leptopion
and leptosigma to $e^+e^-$ pairs.
Ortopositronium anomaly determines the value of $f_{\pi_L}$ and
therefore the value of
leptopion-leptonucleon coupling and the decay rate of leptopion to two
photons.  Various decay widths are in accordance with experimental data
and corrections to electroweak decay rates of neutron and muon are small.
The resonances above $1.6 \ MeV$ are identified as string model satellite
 states of $\sigma_L$ ('radial excitations') \\ b) PCAC
hypothesis and sigma model leads to a general model  for
leptohadron production   in the electromagnetic fields of the colliding
nuclei and production rates for leptopion and other leptohadrons are closely
related to the Fourier transform of the instanton density $\bar{E}\cdot
\bar{B}$ of  the electromagnetic field created by nuclei.
The most probable  source of  anomalous $e^+e^-$ pairs is  the production  of
$\sigma_L\pi_L$ pairs from vacuum  followed by
$\sigma_L\rightarrow e^+e^-$ decay.  New
effects are resonance in photon photon scattering at cm energy equal to
leptopion mass and production of $e_{ex}\bar{e}_{ex}$
($e_{ex}$ is leptobaryon with quantum numbers of electron) and
$e_{ex}\bar{e}$ pairs
in heavy ion collisions.\\
c)   Leptopion exchange gives dominating contribution to $\nu-e$ and
$\bar{\nu}-e$ scattering at low energies and a new method of detecting
solar neutrinos is proposed.

\section{Leptohadron hypothesis}

\subsection{Anomalous $e^+e^-$ pairs in heavy ion collisions}

Heavy ion-collision experiments carried out at the Gesellschaft fur
Schwerionenforschung in Darmstadt, West Germany
\cite{Schweppe,Clemente,Cowan,Tsertos}
have yielded
a rather puzzling set of results. The expectation  was that in heavy
ion collisions in which the combined charge of the two colliding ions exceeds
173, a composite nucleus with $Z > Z_{cr}$ would form and
the probability for spontaneous positron emission would become appreciable.

\vm

Indeed, narrow peaks of widths of rougly 50-70 keV and energies
about 350$\pm$ 50 keV were observed in the positron spectra
but it turned out that the position of peaks seems to be a constant
function of Z rather that vary as $Z^{20}$ as expected and that peaks
are generated also for Z smaller than the critical Z. The collision
energies at which peaks occur lie in the neighbourhodd of 5.7-6
MeV/nucleon. Also it was found that positrons are accompanied by $e^-$-
emission. Data are consistent with the assumption that some structure
at rest in cm is formed and decays subsequently to $e^+e^-$ pair.

\vm

Various theoretical explanations for these peaks have been suggested
\cite{Chodos,Kraus}. For example, lines might be created by pair conversion
 in the
presence of heavy nuclei.  In nuclear physics explanations the lines
are due to some nuclear transition that occurs in the compound nucleus
formed in the collision or in the fragmets formed. The Z-independence
of the peaks seems however to exclude both atomic and nuclear physics
explanations \cite{Chodos}. Elementary particle physics explanations
\cite{Chodos,Kraus}
seem to be excluded already by the fact that several peaks have been
observed in the range
$1.6-1.8  \  MeV$  with widths of order $ 10-10^2  \ keV$. These states
decay  to $e^+e^-$ pairs. There
is evidence for one narrow peak  with width
of order one $keV$ at 1.062 Mev \cite{Chodos}: this state decays to
photon-photon pairs.


\vm

Thus it seems that the structures produced might be composite,
perhaps resonances in $e^+e^-$ system. The difficulty of this explanation
is that conventional QED seems to offer no natural explanation for the
strong force needed to explain the energy scale of the states. One
idea is that the strong electromagnetic fields create a new phase of
QED \cite{Chodos} and that the resonances are analogous to pseudoscalar
mesons appearing as resonances in strongly interacting systems.

\vm

TGD explanation proposed  is based on the following
hypothesis motivated by Topological Geometrodynamics  \cite{TGD,padTGD}.\\
a) Ordinary leptons are not point like particles and can have colored
excitations, which form color singlet bound states. A natural identification
for the primary condensate level is $M_{127}$ so that the mass scale
is of order  one $MeV$  for states containing lowest generation colored
leptons.   \\  b)  The states in question are
leptohadrons that is    color confined states formed
from the colored excitations of $e^+$ and $ e^-$.  $m=1.062 \ MeV$ state is
identified as leptopion $\pi_L$  and $m=1.8 \ MeV$ state turns out to be
identifiable as scalar particle $\sigma_L$ predicted by sigma model providing
a realization of PCAC hypothesis. The remaining resonances can be identified
in string model picture as $J=0$ satellites of $\sigma_L(1.8 \ MeV)$.

\vm

The program of the section is  following:\\
a)   PCAC hypothesis successfull in low energy pion physics is generalized
to
the case of leptopion. Hypothesis allows to deduce the coupling of
leptopion to leptons and leptobaryons in terms of leptobaryon-lepton mixing
angles.  Ortopositronium anomaly allows to
deduce precise value of $f_{\pi_L}$ so that the crucial parameters of the
model are completely fixed.
 The decay rates of leptopion to photon pair and   of leptosigma to
ordinary $e^+e^-$ pairs  are within experimental bounds and corrections
to muon and beta decay rates are small.
A new calculable
resonance contribution to  photon-photon scattering at cm energy equal
to leptopion mass is predicted.
 \\
b)  A model for for the leptohadron and leptopion production is constructed.
The starting point is sigma model providing a realization of PCAC hypothesis.
In an external electromagnetic field leptopion develops a vacuum expectation
value proportional to
 electromagnetic anomaly term  \cite{Iztyk} so that
production amplitude  for leptopion is essentially the Fourier transform of
the scalar product of the electric field of the stationary target nucleus
with the magnetic field of the colliding nucleus.
Sigma model makes it possible to relate the production amplitude for
$\sigma_L \pi_L$ pairs to  the leptopion production amplitude: the key
element of the model is the large value of the $\sigma\pi_L\pi_L$ coupling
constant.  The decays of the  scalar particle $\sigma_L$ ($1.8 \ MeV$ state)
and its radial excitations directly to  $e^+e^-$ states   is the simplest
explanation for the production  of $e^+e^-$ pairs. The fact that two-particle
states are produced could perhaps   explain the observed  deviations from the
simple resonance decay picture.  Model predicts also a direct production of
leptobaryon ($e_{ex}$) pairs in heavy ion collisions.
 \\
c) Leptohadron production amplitudes  are proportional to
leptopion production amplitude and this motivates a detailed study of
leptopion production.  Two models for leptopion
production are developed: in  classical  model
 colliding nucleus
is treated classically whereas in  quantum
 model the colliding nucleus
is described quantum mechanically.  It turns out
that classical model explains the peculiar production characteristics
of leptopion but that production cross section is too small by several
orders of magnitude.
Quantum mechanical model  predicts also diffractive effects: production
cross section varies rapidly as a function of the scattering angle and for a
fixed value of scattering angle there is a rapid variation with the
collision velocity. The estimate for the total $e^+e^-$ production cross
 section
is of correct order of magnitude due to the coherent summation of the
contributions to the amplitude from different values of the  impact parameter
 at the peak.
 \\
c) The problem of understanding  the effective  experimental
absence of  leptohadronic color
interactions in TGD picture is discussed.

\subsection{Leptopions and generalized PCAC hypothesis}

One can say that the PCAC hypothesis predicts the existence of pions and a
connection between the pion nucleon coupling strength and the pion decay rate
to leptons. In the following we give the PCAC argument and its
generalization and consider various consequences.

\subsubsection{PCAC for ordinary pions}

The PCAC argument for ordinary pions goes as follows \cite{Okun}:\\
a) Consider the contribution of the hadronic axial current to the matrix
element describing lepton nucleon scattering (say $N+\nu \rightarrow P+e^-$)
by weak interactions. The contribution in question reduces to the well-known
current-current form

\begin{eqnarray}
M&=& \frac{G_F}{\sqrt{2}}g_A L_{\alpha}\langle P\vert A^{\alpha}\vert
P\rangle\nonumber\\
L_{\alpha}&=& \bar{e}\gamma_{\alpha}(1+\gamma_5)\nu\nonumber\\
\langle P\vert A^{\alpha}\vert
P\rangle &=& \bar{P}\gamma^{\alpha}N
\end{eqnarray}

\noindent where $G_F=\frac{ \pi \alpha}{2m_W^2 sin^2(\theta_W)} \simeq
10^{-5}/m_p^2$
 denotes the
dimensional weak interaction coupling strength and $g_A$ is the nucleon axial
form factor:$g_A\simeq 1.253$.\\
b)  The matrix element of the hadronic axial current is not divergenceless,
due to the nonvanishing nucleon mass,

\begin{eqnarray}
a_{\alpha}\langle P\vert A^{\alpha}\vert
P\rangle&\simeq &2m_p \bar{P}\gamma_5 N
\end{eqnarray}

\noindent Here $q^{\alpha}$ denotes the momentum transfer vector. In order
to obtain divergenceless current, one can modify the expression for the
matrix element of the axial current

\begin{eqnarray}
\langle P\vert A^{\alpha}\vert N\rangle& \rightarrow& \langle P\vert
A^{\alpha} \vert N\rangle - q^{\alpha}2m_p  \bar{P}\gamma_5 N \frac{1}{q^2}
\end{eqnarray}

\noindent c) The modification introduces a new term to the lepton-hadron
scattering amplitude identifiable as an exchange of a massless pseudoscalar
particle

\begin{eqnarray}
\delta T &=& \frac{G_Fg_A}{\sqrt{2}} L_{\alpha}\frac{2m_p q^{\alpha}}{q^2}
\bar{P}\gamma_5N
\end{eqnarray}

\noindent  The amplitude is identifiable as the amplitude describing the
exchange of the pion, which gets its mass via the breaking of chiral
invariance and one obtains by the straightfowread replacement $q^2\rightarrow
q^2-m_{\pi}^2$ the correct form of the amplitude.\\
d) The nontrivial point is that the interpretations as pion exhange is indeed
possible since the amplitude obtained is to a good approximation identical
to that obtained from the Feynman diagram describing pion exchange, where
the pion nucleon coupling constant and pion decay amplitude appear

\begin{eqnarray}
T_2&=& \frac{G}{\sqrt{2}}f_{\pi}q^{\alpha}L_{\alpha} \frac{1}{q^2-m_{\pi}^2}
g\sqrt{2}\bar{P}\gamma_5 N
 \end{eqnarray}

\noindent The condition $\delta T\sim T_2$  gives from Goldberger-Treiman
 \cite{Okun}

\begin{eqnarray}
g_A (\simeq 1.25) &=& \sqrt{2}\frac{f_{\pi}g}{2m_p} (\simeq 1.3)
\end{eqnarray}

\noindent satisfied in a  good accuracy experimentally.

\subsubsection{PCAC in leptonic sector}

A natural question is why not generalize the previous argument to the
leptonic sector and look at what one obtains. The generalization is based
on following general picture.\\
a) There are two levels to be considered: the level of ordinary leptons and
the level of leptobaryons  of type  $f_{ABC}L_{8}^AL_{8}^B\bar{L}_{8}^C$
possessing same quantum numbers as leptons.  The interaction transforming
these states to each other causes in mass eigenstates  mixing of leptobaryons
 with ordinary leptons described by mixing angles. The masses
of lepton and corresponding leptobaryon could be quite near to each other
and in case of electron should be the case as it turns out.  \\
b) A counterargument against the applications of PCAC hypothesis
at level of ordinary leptons  is that baryons and mesons
are both bound states of quarks whereas   ordinary leptons
are not bound states of color octet leptons. The divergence of  the
axial current  is however completely independent of the possible internal
 structure of leptons and microscopic emission mechanism.
Ordinary lepton
cannot emit leptopion directly but must first transform to leptobaryon
with same quantum numbers:
 phenomenologically this process can be described using  mixing angle
$sin(\theta_B)$.
The  emission of leptopion
proceeds  as  $L\rightarrow B_L \ $: $B_L \rightarrow B_L +\pi_L$:
$B_L\rightarrow L $, where $B_L$ denotes leptobaryon of type
 structure $f_{ABC}L_{8}^AL_{8}^B\bar{L}_{8}^C$. The transformation amplitude
 $L\rightarrow B_L$ is
 proportional to  the mixing angle $sin(\theta_L)$.

\vm

 Three different  PCAC type  identities
are assumed to hold true:

\vm

\noindent PCAC1) The vertex for the emission of leptopion by ordinary lepton
 is equivalent
with the graph in which lepton  $L$ transforms to
leptobaryon  $L^{ex}$ with same quantum numbers,  emits leptopion and
transforms back to ordinary lepton. The assumption relates the couplings
$g(L_1,L_2)$ and $g(L^{ex}_1,L^{ex}_2)$ (analogous to strong coupling) and
mixing angles to each other

\begin{eqnarray}
g(L_1,L_2)&=& g(L^{ex}_1,L^{ex}_2)sin(\theta_1)sin(\theta_2)
\end{eqnarray}

\noindent The condition implies that in electroweak interactions ordinary
leptons do not transform to their exotic counterparts.

\vm

 \noindent PCAC2) The generalization of  the ordinary Goldberger-Treiman
argument holds true, when ordinary baryons are replaced with leptobaryons.
This gives the condition expressing the coupling $f(\pi_L)$
 of the  leptopion state to axial current defined as

\begin{eqnarray}
\langle vac \vert A_{\alpha}\vert \pi_L\rangle = ip_{\alpha}f_{\pi_L}
 \end{eqnarray}

\noindent in terms of the masses of
leptobaryons and strong coupling $g$.

\begin{eqnarray}
f_{\pi_L}&=& \sqrt{2}g_A\frac{(m_{ex}(1)+ m_{ex}(2)) sin(\theta_1)
sin(\theta_2)} {g (L_1,L_2)}
\end{eqnarray}

\noindent  where $g_A$ is  parameter characterizing
the deviation  of  weak coupling strength associated with
leptobaryon from ideal value:    $g_A\sim 1$ holds true in good approximation.

\vm

\noindent  PCAC3)  The elimination of leptonic axial anomaly from leptonic
current fixes the values of $g(L_i,L_j)$. \\
i)   The standard contribution
to the scattering of leptons by weak interactions.sis
given by the expression

\begin{eqnarray}
T&=& \frac{G_F}{\sqrt{2}}\langle L_1 \vert A^{\alpha} \vert L_2\rangle
\langle L_3\vert A_{\alpha} \vert L_4\rangle\nonumber\\
\langle L_i \vert A^{\alpha} \vert L_j\rangle&=& \bar{L}_i
\gamma^{\alpha}\gamma_5 L_j
\end{eqnarray}

\noindent ii)  The elimination of the leptonic axial anomaly

\begin{eqnarray}
q_{\alpha}\langle L_i\vert A^{\alpha}\vert
L_j\rangle&= &(m(L_i)+m(L_j) )\bar{L}_i\gamma_5L_j
\end{eqnarray}

\noindent by modifying the axial current by the  anomaly term

\begin{eqnarray}
\langle L_i\vert A^{\alpha}\vert L_j\rangle& \rightarrow& \langle L_i\vert
A^{\alpha} \vert L_j\rangle -(m(L_i)+m(L_j))\frac{q^{\alpha}}{q^2}
\bar{L_i}\gamma_5 L_j
  \end{eqnarray}

\noindent induces a new interaction term in the scattering
of ordinary leptons.\\
 iii) It is assumed that this term is equivalent with the
exchange of leptopion.   This fixes the value of the coupling constant
$g(L_1, L_2)$ to

\begin{eqnarray}
g(L_1,L_2) &=& 2^{1/4}\sqrt{G_F}(m(L_1)+m(L_2))\xi\nonumber\\
\xi (charged) &=& 1\nonumber\\
\xi (neutral)&= &cos(\theta_W)
\label{couplings}
\end{eqnarray}

\noindent Here the coefficient $\xi$ is related
to different values of masses for gauge bosons $W$ and $Z$ appearing
in charged and neutral current interactions. An  important factor $2$ comes
from the modification of the axial current in  both matrix elements of
the axial current.

\vm

 Leptopion exchange  interaction couples right and  left handed leptons to
each other and its strength is of the same order of magnitude as the strength
of the ordinary weak interaction at energies not considerably large than the
mass of the leptopion.  At high energies this interaction is negligible and
the existence of the leptopion predicts no corrections to the parameters  of
the standard model since these are determined from weak interactions at much
higher energies. If leptopion mass is sufficiently small (as found,
$m(\pi_L)<2m_e$ is allowed by the experimental data), the interaction mediated
by leptopion exchange can become quite strong due to the presence of the
leptopion progator.
 The value  of the lepton leptopion coupling  is $g(e,e)\equiv g\sim 5.6
\cdot 10^{-6}$.  It is perhaps worth noticing that the value of the coupling
constant is of the same order as lepton-Higgs coupling constant and also
proportional to the mass of the lepton. This is accordance with the idea
that the  components of the Higgs boson correspond to the divergences of
various vector currents \cite{TGD}. What is important that the value of the
coupling is completely independent of the details of leptopion emission.
\vm

PCAC  identities fix the values of coupling constants apart from the values
of mixing angles.
If one assumes that the strong interaction mediated by leptopions is
really strong and the coupling strength $g(L_{ex},L_{ex})$ is of same
order of magnitude as the ordinary pion nucleon coupling strength $g(\pi
NN)\simeq 13.5$ one obtains an
estimate for the value of the mixing angle $sin(\theta_e)$\\
$sin(\theta_e)\sim \frac{g(\pi NN)}{g(L,L)} \sim 2.4 \cdot 10^{-6}$.
 This implies the order of magnitude $10^{-11}m_W\sim eV$
 for $f_{\pi_L}$.
The experimental bound for $\pi_L\rightarrow \gamma \gamma$ decay width
 implies that $sin(\theta_e)$ must be at least by a factor
$10^{5/4}$ larger.Ortopositronium decay rate anomaly $\Delta \Gamma/\Gamma
\sim 10^{-3}$  and the assumption  $m_{ex}\geq 1.3 \ MeV$ (so
that $e_{ex}\bar{e}$ decay is not
possible)  gives the upper bound
 $sin(\theta_e)\leq x \cdot 10^{-4}$, where
the value of $x \sim 1$ depends on the number
of leptopion type states and on the precise value of Op anomaly.
 Leptohadronic strong coupling satisfies $g\leq g_{max}\sim 1$.

\subsection{Leptopion decays and  PCAC hypothesis}

The PCAC argument makes it possible to predict the leptopion coupling and
decay rates of leptopion  to various channels.  Actually the orders of
magnitude for the  decay rates of leptosigma and other leptomesons can be
deduced also. The comparison with the
experimental data is made difficult by the uncertainty of
the identifications.  The lightest
candidate has mass $1.062 \ MeV$ and decay width of order 1 keV \cite{Chodos}:
only photon photon decay has been observed for this state. The
next leptomeson candidates are in the  mass range $1.6-1.8 \ MeV$.
Perhaps the best status is possessed by 'Darmstadtium'  with  mass
$1.8 \ MeV$. For these
states  decays
 final  states identified as $e^+e^-$ pairs dominate: these states
probably correspond to  leptosigma and its string model satellites
('radial excitations').
 Hadron physics experience
 suggests that
the decay widths of the  leptohadrons (leptopion forming a possible
exception)
 should be about  1-10 per cent of particle mass
 as in
hadron physics.  The upper bounds for
the widths are indeed in the range $50-70  \ keV$ \cite{Chodos}.

\vm

a)   As in the case of the ordinary pion,  anomaly considerations give the
following  approximate expression for the decay rate of the leptopion
 to two-photon final
states \cite{Iztyk})

\begin{eqnarray}
\Gamma (\pi_L \rightarrow \gamma\gamma) &=& \frac{\alpha^2
m^3(\pi_L)}{64f_{\pi_L}^2 \pi^3} = (\frac{m(\pi_L)}{m(\pi)})^3
\frac{f^2_{\pi}}{f^2_{\pi_L}} \Gamma (\pi \rightarrow \gamma \gamma )
\end{eqnarray}

\noindent where the decay rate for the ordinary pion is given by
$\Gamma (\pi
\rightarrow \gamma \gamma )\simeq 1.5 \cdot 10^{-16} \ sec$.
For $m(\pi_L)= 1.062 \ MeV$ and $f_{\pi_L}= 7.9 \ keV$ implied by the
ortopositronium decay rate anomaly $\Delta \Gamma/\Gamma=10^{-3}$ one has
$\Gamma(\gamma \gamma)= .52 \ keV$, which is  consistent with
 the experimental estimate of order $1 \ keV$ \cite{Chodos}.
  Actually
   several leptopion states (string model satellites) could exist
 in one-one correspondence with $\sigma$ scalars
 (3 at least). Since all 3 leptopion states contribute to Op decay
rate, the actual value of $f_{\pi_L}$  assumed to  scale as $m(\pi_L)$, is
actually larger in this case: it turns out that $f_{\pi_L}$
for the lightest leptopion increases
to $f_{\pi_L}= 11 \ keV$  and gives $\Gamma(\gamma \gamma)\simeq  .27 \ keV$.
The increase of the ortopositronium anomaly by a factor of, say $4$,
 implies
corresponding decrease in $f_{\pi_L}^2$.  The value of $f_{\pi_L}$ is
also sensitive to the precise  value of the mass of the lightest leptopion.
The production cross section for anomalous $e^+e^-$ pairs turns out
to be proportional to $f_{\pi_L}^{-4}$ and is very sensitive to the
exact value of $f_{\pi_L}$: $f_{\pi_L}\sim 1 \ keV$ is favoured and
corresponds to Op anomaly $\Delta \Gamma/\Gamma\sim 4\cdot 10^{-3}$ in
3-leptopion case.

\vm

b) The value of the leptopion-lepton coupling can be used to predict the
 decay rate of
leptopion to leptons. One obtains for the decay rate
$\pi^0_L \rightarrow e^+e^-$ the estimate

\begin{eqnarray}
\Gamma (\pi_L\rightarrow e^+e^-) &=&
4\frac{g(e,e)^2\pi}{2(2\pi)^2}(1-4x^2) m(\pi_L)
\nonumber\\
&=& 16Gm_e^2cos^2(\theta_W)
\frac{\sqrt{2}}{4\pi}(1-4x^2)  m(\pi_L)\nonumber\\
x&=& \frac{m_e}{m(\pi_L)}
\end{eqnarray}

\noindent for the decay rate of the leptopion:  for  leptopion
mass $m(\pi_L)\simeq 1.062 \ MeV$ one obtains for the decay rate  the
estimate  $\Gamma \sim 1/( 1.3\cdot  10^{-8}  \ sec)$: the low decay rate is
partly   due to the
phase space suppresion and implies that $e^+e^-$ decay products
cannot be observed in the  measurement volume. The low decay rate is in
 accordance
with the identification of the leptopion as the $m=1.062 \ MeV$ leptopion
candidate. In sigma model leptopion and leptosigma  have identical
lifetimes and for   leptosigma mass of order $1.8 \ MeV$ one
obtains $\Gamma (\sigma_L \rightarrow e^+ e^-)\simeq 1/( 8.2\cdot 10^{-10} \
sec)$: the prediction  is larger
than   the lower limit $ \sim  1/(10^{-9} \ sec )$ for the decay  rate
 implied by the
 requirement that $\sigma_L$ decays inside the measurement volume.
  The estimates of the lifetime obtained from
heavy ion collisions \cite{Koenig} give  the estimate  $\tau \geq 10^{-10} \
sec$.  The large value of the lifetime is
 in accordance with the limits for the lifetime obtained from Babbha
scattering \cite{Judge}, which indicate that the  lifetime must be longer
than $10^{-12}$ sec.

\vm

 For leptomeson candidates  with mass above $1.6 \ MeV$
no experimental evidence for
other decay modes than $X \rightarrow e^+e^-$
 has been found and the
empirical upper limit for $\gamma\gamma/e^+e^-$ branching ratio
\cite{Dantzman} is $\Gamma (\gamma\gamma)/\Gamma
( e^+e^-)
\leq  10^{-3}$.  If the identification of  the decay products
as $e^+e^-$ pairs  (rather than  $e_{ex} \bar{e}_{ex}$
pairs of leptonucleons!)  is correct then the only possible conclusion is
that these states cannot correspond to leptopion since leptopion should
decay dominantly into photon photon pairs.

\vm

c)  The expression for the decay rate $\pi_L\rightarrow e+\bar{\nu_e}$
reads as

\begin{eqnarray}
\Gamma (\pi_L^- \rightarrow e\nu_e) &=&
8Gm_e^2\frac{(1-x^2)^2}{2(1+x^2)} \frac{\sqrt{2}}{(2\pi)^5} m(\pi_L)
\nonumber\\
&=& \frac{4}{cos^2(\theta_W)}\frac{(1-x^2)}{(1+x^2)(1-4x^2)}
\Gamma (\pi_L^0\rightarrow e^+e^-)
 \end{eqnarray}

\noindent and gives   $\Gamma (\pi_L^- \rightarrow e\nu_e) \simeq
1/( 3.6 \cdot 10^{-10} \ sec)$    for $m(\pi_L)=1.062 \ MeV$.

\vm

d) One must consider also  the
possibility that leptopion decay products are either $e_{ex} \bar{e}_{ex}$
or $e_{ex}\bar{e}$ pairs with $e_{ex}$ having mass of near the mass of
electron so that it could be misidentified as electron although
  the experience with the
ordinary hadron physics does not give support to this possibility.
  If
the mass of leptonucleon $e_{ex}$ with quantum numbers of electron is smaller
than $m(\pi_L)/2$  it can be produced in leptopion annihilation.
One must also assume $m(e_{ex})>m_e$: otherwise electrons would
spontaneously decay to leptonucleons via photon emission. The production rate
to   leptonucleon pair  can be written as

\begin{eqnarray}
\Gamma (\pi_L\rightarrow e^+_{ex}e^-_{ex}) &=&
\frac{1}{sin^4(\theta_e)}\frac{(1-4y^2)}{(1-4x^2)}
\Gamma (\pi_L\rightarrow e^+e^-) \nonumber\\
y&=& \frac{m(e_{ex})}{m(\pi_L)}
\end{eqnarray}

\noindent If $e-e_{ex}$ mass difference is sufficiently small the kinematic
suppression does not differ significantly from that for $e^+e^-$
 pair.
The limits from Babbha scattering give no bounds on the rate
of $\pi_L \rightarrow e^+_{ex}e^-_{ex}$ decay.   The  decay
rate $ \Gamma \sim 10^{26}/ sec$ implied by $sin(\theta_e)  \sim
10^{-4}$ implies decay width of order  one $10  GeV$, which does not make
sense so that
the  constraint $me(e_{ex})> m(\pi_L)/2$ follows. The same
argument applied to $1.8 \ MeV$ states implies the lower
 bound $m(e_{ex})>.9 \
MeV$.

\vm

e) The decay rate of leptosigma to $\bar{e}e_{ex}$ pair has  sensible
order of magnitude: for $sin(\theta_e) =1.2 \cdot 10^{-4}$, $ m_{\sigma_L}=
1.8 \ MeV$ and $m_{e_{ex}}=1.3 \ MeV$ one has
 $\Gamma \simeq  6\ keV$  allowed by the experimental
limits.
 This decay
is kinematically possible only provided the mass of  $e_{ex}$ is
in below  $1.3 \ MeV$.
    These decays should dominate
by a factor $1/sin^2(\theta_e)$ over $e^+e^-$ decays
if kinematically allowed.
 A signature of these events, if identified erraneously as
electron positron pairs,  is
the nonvanishing value of the energy difference in the cm frame of the  pair:
$E(e^-)-E(e^+)\simeq (m^2(e_{ex})-m_e^2)/2E> 160 \ keV$ for $E=1.8 \ MeV$.
If the  decay  $e_{ex} \rightarrow e+\gamma$ takes place
before the  detection the energy asymmetry changes its sign.
Energy asymmetry
Ê\cite{Asymmetry}  increasing
 with the rest energy of the decaying object has indeed been observed:
 the proposed interpretation has been that
 electron forms a  bound
state with the second nucleus so that its energy is lowered.
Also a deviation from the
momentum distribution implied by the decay of neutral particle
to $e^+e^-$ pair (momenta are opposite in the rest frame)
results from the emission of photon. This kind of deviation has also
been observed \cite{Asymmetry}: the  proposed explanation is that third
object is involved in the decay.  A possible alternative explanation for the
asymmetries is the production mechanism ($\sigma_L\pi_L$ pairs instead of
single particle states).

\vm

 f)  The decay    to electron and photon would
be  a unique signature of $e_{ex}$.   The general feature of
of fermion family mixing is  that mixing takes place
in charged currents. In present case mixing is of different type
 so that
$e_{ex}\rightarrow e+\gamma$  might  be allowed. If this is not the
case then the decay takes place as weak decay via the emission of virtual
$W$ boson: $e_{ex}\rightarrow e+\nu_{e}+\bar{\nu}_e$ and is very slow
 due to the presence of mixing angle and kinematical supression.
The energy of the emitted
photon  is
 $E_{\gamma}= (m_{ex}^2-m_e^2)/2m_e $.
The decay rate $\Gamma(e_{ex}\rightarrow e+\gamma)$ is given by

\begin{eqnarray}
\Gamma (e_{ex}\rightarrow e+\gamma)&=& \alpha_{em}sin^2(\theta_e)Xm_e
\nonumber\\
X&=& \frac{(m_1-m_e)^3(m_1+m_e) m_e}{(m_1^2+m_e^2)^2m_1}\nonumber\\
\
\end{eqnarray}

\noindent For $m(e_{ex})=1.3 \ MeV$ the decay  of order
$1/(1.4 \cdot  10^{-10} \ sec)$  for $sin(\theta_e)= 1.2\cdot 10^{-4}$
so that   a considerable fraction of leptonucleons would decay to electrons in
the measurement volume. In the experiments
positrons are identified via pair annihilation and since pair annihilation
rate for $\bar{e}_{ex}$ is by a factor $sin^2(\theta_e)$ slower than for
$e^+$ the  particles identified as positrons must indeed be positrons.
For
sufficiently  small
mass difference $m(e_{ex})-m_e$ the particles identified as electron could
actually be $e_{ex}$. The decay of $e_{ex}$ to electron plus photon before
its detection
seems however more reasonable alternative since it
could explain the observed energy asymmetry \cite{Asymmetry}.

\vm

g)  The results
have several implications as far as the decays of on mass shell states are
considered: \\
i)  For $m(e_{ex})>1.3 \ MeV$ the only kinematically possible  decay
mode is the decay to $e^+e^-$ pair. Production mechanism might
explain the asymmetries \cite{Asymmetry}. The decay rate
of on mass shell $\pi_L$  and $\sigma_L$ (or  $\eta_L,\rho_L,..$) is  above
the  lower limit   allowed by the  detection in the   measurement
 volume. \\
ii) If the mass of $e_{ex}$ is larger than $.9 \ MeV$ but smaller than $1.3 \
MeV$ $e_{ex}\bar{e}$ decays dominate over $e^+e^-$ decays. The decay
 $e_{ex}\rightarrow e+\gamma$  before detection could explain the observed
energy asymmetry.
 \\
iii) It will be found that the direct production of $e_{ex}\bar{e}$ pairs is
also possible in  the heavy ion collision but the
rate is much smaller due to the smaller phase space volume in two-particle
case. The annihilation rate of
 $\bar{e}_{ex}$ in matter is by a factor
$sin^2(\theta_e)$  smaller than the annihilation rate of positron. This
provides a unique signature of $e_{ex}$ if
 $e^+$  annihilation rate in matter is larger than the decay rate of
$\bar{e}_{ex}$. In lead the lifetime of positron is $\tau \sim 10^{-10} \ sec$
and indeed larger that $e_{ex}$ lifetime.


\vm

h) A brief comment on the Karmen anomaly  \cite{Karmen} observed
in the decays of $\pi^+$ is
 in order.  The anomaly suggests the existence  \cite{Karmen1} of new
weakly interacting neutral particle $x$, which mixes with muon neutrino.  One
class
of solutions to laboratory constraints,  which might  evade also
cosmological and astrophysical constraints,  corresponds to
object  $x$ mixing with muon type neutrino and decaying radiatively
to $\gamma +\nu_{\mu}$ via the emission of virtual $W$ boson. The value of
the mixing parameter $U(\mu,x)$ describing $\nu_{mu}-x$ mixing
satisfies $\vert U_{\mu, x}\vert^4\simeq .8 \cdot 10^{-10}$.
 The following  naive PCAC argument gives order of magnitude
estimate for
$\vert U(\mu,x)\vert \sim
sin(\theta_{\mu})$. The value of  $g(\mu,\mu)$
is by a factor $m_{\mu}/m_e$ larger than $g(e,e)$. If the leptohadronic
couplings  $g(\mu_{ex},\mu_{ex}) $ and $g(e_{ex}, e_{ex})$ are of same order
of magnitude then one has
 $sin(\theta_{\mu})\leq .02$ (3 leptopion states and Op anomaly equal to
$Op=5\cdot 10^{-3}$): the lower bound  is
  $6.5$
times larger than
the value  $.003$ deduced in \cite{Karmen1}.  The actual value could be
 considerably smaller since $e_{ex}$ mass could be larger than $1.3 \ MeV$ by
a factor of order $10$.

\subsection{Leptopions and weak decays }

The couplings of leptomeson to electroweak gauge bosons can be  estimatd
using PCAC and CVC hypothesis \cite{Iztyk}.   The effective
$m_{\pi_L}-W$ vertex is  the matrix element of electroweak axial current
between vacuum and charged leptomeson state
 and can be deduced using same arguments as in the case of ordinary
charged pion

\begin{eqnarray}
 \langle 0 \vert J^{\alpha}_A \vert \pi_L^-\rangle= K m(\pi_l)p^{\alpha}
\nonumber\\
\end{eqnarray}

\noindent  where $K$ is some numerical factor and $p^{\alpha}$ denotes the
momentum of leptopion.
 For neutral leptopion the same argument gives vanishing coupling to photon
by the conservation of vector current. This has the important consequence
that  leptopion cannot be observed  as  resonance in $e^+e^-$
annihilation  in single photon channel.  In two photon channel leptopion
should appear as resonance.  The
 effective interaction Lagrangian is the 'instanton' density of
electromagnetic field \cite{Heavy,Lepto}  giving additional contribution
to the divergence of the
 axial current and was used to derive a model for leptopion production  in
heavy ion collisions.

\subsubsection{ Leptohadrons and lepton decays}

  The  lifetime of charged leptopion   is  from PCAC estimates larger
 than   $10^{-10}$ seconds by the previous
PCAC estimates.   Therefore leptopions are practically stable particles and
can appear in
the final states of particle reactions.  In particular, leptopion atoms
are possible and  by Bose statistics have the peculiar property that
ground state can contain many leptopions.

\vm

 Lepton decays $L \rightarrow \nu_{\mu}+ H_L$, $L=e,\mu,\tau$  via
emission of virtual $W$ are kinematically  allowed and an anomalous
resonance  peak in the neutrino energy  spectrum at  energy

\begin{eqnarray}
E(\nu_{L})&=&\frac{m(L)}{2} -\frac{m_H^2}{2m(L)}
\end{eqnarray}

\noindent provides a  unique test for the leptohadron hypothesis.  If
 leptopion is too light   electrons  would decay to charged leptopions and
neutrinos unless the condition $m(\pi_L)>m_e$  holds true.

\vm

   The existence of  a  new  decay channel for muon  is an obvious danger
to the leptohadron scenario:
 large changes in muon decay rate are not allowed. \\
  a)   Consider first the decay  $\mu\rightarrow \nu_{\mu}+\pi_L$ where
$\pi_L$ is  on mass  shell leptopion.
  Leptopion has energy $\sim m(\mu)/2$ in muon rest system  and is highly
relativistic so that in  the muon rest system the
 lifetime  of leptopion is
 of order $\frac{m(\mu)}{2m(\pi_L)}\tau(\pi_L) $  and the average length
traveled
 by leptopion before decay is of order $10^8$ meters! This means that
leptopion
 can be treated as stable particle.  The presence of a new decay channel
changes the lifetime of muon although the  rate for events using $e\nu_e$
pair as  signature is not changed.  The effective  $H_L-W$ vertex was
deduced above.
 The rate for the decay via leptopion emission and its ratio to ordinary
rate
 for muon decay are given by

\begin{eqnarray}
\Gamma (\mu \rightarrow \nu_{\mu} + H_L)&=& \frac{G^2K^2}{2^5
\pi}m^4(\mu)m^2(H_L)(1-\frac{m^2(H_L)}{m^2(\mu)})
\frac{(m^2(\mu)-m^2(H_L))}{(m^2(\mu)+m^2(H_L))   }\nonumber\\
\frac{\Gamma (\mu \rightarrow \nu_{\mu} + H_L)}{\Gamma (\mu \rightarrow
 \nu_{\mu}+e+\bar{\nu}_e)} &=& 6 \cdot (2\pi^4) K^2
\frac{m^2(H_L)}{m^2(\mu)}\frac{(m^2(\mu)-m^2(H_L))}{(m^2(\mu)+m^2(H_L))   }
\nonumber\\
\
\end{eqnarray}

  \noindent and is of order $.93  K^2$ in case of leptopion.  As far as
the determination of $G_F$  or equivalently $m_{W}^2$ from muon decay rate
is considered the situation seems to be good since  the change
 introduced to $G_F$ is of order $\Delta G_F/G_F \simeq  0.93 K^2$ so that
$K$ must be considerably smaller than one.
 For the physical value of $K$:  $K\leq 10^{-2}$ the contribution to the muon
 decay
rate is neglibigle

\vm

  Leptohadrons can appear also as virtual particles  in the decay
amplitude $\mu\rightarrow \nu_{\mu}+e\nu_e$ and this changes the value of
muon decay rate. The correction is however extremely small since the decay
vertex of  intermediate off mass shell leptopion is proportional to its
decay rate.

\subsubsection{ Leptopions and beta decay}

 If leptopions are allowed as final state particles leptopion emission
provides a new   channel  $n\rightarrow p+ \pi_L$ for  beta  decay of
nuclei since the invariant mass of virtual  $W$ boson varies within the
 range $(m_e=0.511 \ MeV,m_n-m_p=1.293 MeV$.  The resonance peak for
$m(\pi_L) \simeq 1 \ MeV$ is extremely sharp due to the long lifetime of
the charged leptopion. The energy of  the leptopion at resonance is

\begin{eqnarray}
E(\pi_L)&=& (m_n-m_p)\frac{(m_n+m_p)}{2m_n}+\frac{m(\pi_L)^2}{2m_n}
\simeq m_n-m_p
\end{eqnarray}

\noindent Together with long lifetime this
 leptopions escape the detector volume without decaying (the exact
knowledge of the energy
 of charged leptopion  might make possible  its direct detection).

\vm

 The contribution of leptopion to neutron decay rate is not negligible.
Decay  amplitude is proportional to
 superposition of divergences of axial and vector currents between proton
and neutron states.

\begin{eqnarray}
M&=&  \frac{G}{\sqrt{2}} Km(\pi_L) ( q^{\alpha}V_{\alpha}+
 q^{\alpha}A_{\alpha})
\end{eqnarray}

\noindent    For  exactly conserved vector current the contribution of
vector current vanishes identically. The matrix element of the  divergence
of axial vector current at small momentum transfer (approximately zero)
is in good approximation given by

\begin{eqnarray}
\langle p \vert q^{\alpha}A_{\alpha}  \vert n\rangle&=& g_A(m_p+m_n)
\bar{u}_p\gamma_5u_n\nonumber\\
g_A&\simeq& 1.253
\end{eqnarray}

\noindent
 Straightforward calculation shows that the ratio for the decay rate  via
leptopion emission and ordinary beta decay rate is in good approximation
 given by

\begin{eqnarray}
\frac{\Gamma(n\rightarrow p+\pi_L)}{\Gamma(n\rightarrow
p+e+\bar{\nu}_e)}  &=&
\frac{30\pi^2g_A^2K^2}{0.47 \cdot (1+3g_A^2)  }
\frac{m_{\pi_L}^2(\Delta^2-m_{\pi_L}^2) ^2   }
{\Delta^6}
\nonumber\\
\Delta&=& m(n)-m(p)
\end{eqnarray}

\noindent Leptopion contribution is smaller than ordinary contribution
if the condition

\begin{eqnarray}
K&<& (\frac{.47\cdot (1+3g_A^2)}{30\pi^2g_A^2}
\frac{ \Delta^6}
{ (\Delta^2-m_{\pi_L}^2) ^2m_{\pi_L}^2} )^{1/2}
\simeq .28
\end{eqnarray}

\noindent  is satisfied. The upper bound $K\leq  10^{-2}$ coming from
the leptopion decay
width and Op anomaly implies that the contribution of leptopion to beta decay
rate is very small.

\subsection{Ortopositronium puzzle and leptopion  in photon photon
scattering}

The decay rate of ortopositronium (Op) has been found to be slightly larger
than the rate predicted by QED \cite{orto,orto1}: the discrepancy is of order
$\Delta \Gamma/\Gamma \sim 10^{-3}$. For parapositronium no anomaly has been
observed. Most of the proposed explanations  \cite{orto1} are based on the
decay mode $Op \rightarrow X+ \gamma$,
where $X$ is some exotic particle. The experimental limits on the branching
ratio $\Gamma(Op\rightarrow X+\gamma)$ are below the required value of order
$10^{-3}$. This  explanation  is excluded also by the standard cosmology
\cite{orto1}.

\vm

 Leptopion hypothesis suggests an obvious solution of  the Op-puzzle.
The increase in annihilation rate is due to the additional contribution
to $Op\rightarrow 3\gamma$  decay coming from the decay  $Op\rightarrow
\gamma_V$ ($V$ denotes 'virtual') followed by the decay $\gamma_V
\rightarrow \gamma+\pi_L^V$  followed by the decay  $\pi_L^V \rightarrow
 \gamma+\gamma$ of the virtual
leptopion to two photon state.   $\gamma \gamma \pi_L$ vertices are induced
by the axial current anomaly $\propto  E\cdot B$. Also
 a modification of parapositronium decay rate is  predicted.
The first contribution comes
 from the decay $Op\rightarrow \pi_L^V\rightarrow \gamma+\gamma$ but
the contribution is very small due the smallness of the coupling $g(e,e)$.
The second contribution  obtained from ortopositronium contribution
by replacing  one outgoing photon with a loop photon is also small.
Since the production of real leptopion is impossible  the mechanism is
consistent with experimental constraints.

\vm

The modification to the  Op annihilation amplitude comes in a good
approximation  from  the  interference term between the  ordinary $e^+e^-$
annihilation  amplitude $F_{st}$
and leptopion induced annihilation amplitude $F_{new}$:

\begin{eqnarray}
\Delta \Gamma &\propto& 2Re( F_{st}\bar{F}_{new})
\end{eqnarray}

\noindent and rough order of magnitude estimate suggests     $\Delta
\Gamma/\Gamma \sim K^2/e^2 = \alpha^2/4\pi\sim 10^{-3}$. It turns out that
the sign  and the order of magnitude of the new
contribution are correct for $f_{\pi_L}\sim 2 \ keV$ deduced also from
the anomalous  $e^+e^-$ production rate.

\vm

  The  new contribution to  $e^+e^-\rightarrow 3\gamma$ decay
amplitude is most easily derivable using for leptopion-photon interaction the
effective action

\begin{eqnarray}
L_1&=& K\pi_L F\wedge F\nonumber \\
K&=& \frac{\alpha_{em}}{8\pi f_{\pi_L}}
\end{eqnarray}

\noindent where $F$ is quantized electromagnetic field.
  The
calculation of the leptopion contribution proceeds in manner  described in
\cite{Iztyk}, where the expression for the  standard contribution and an
elegant  method for treating the average over $e^+e^-$ spin triplet states
and sum over photon polarizations can be found.
 The contribution to
decay rate can be written as

\begin{eqnarray}
\frac{\Delta \Gamma}{\Gamma}&\simeq& K_1 I_0\nonumber\\
K_1&=& \frac{3\alpha}{(\pi^2-9)2^{9}(2\pi)^3}  (\frac{m_e}{f_{\pi_L}})^2
\nonumber\\ I_0&=& \int_0^{1}\int_{-1}^{umax} \frac{f}{v+f-1-x^2}  v^2 (
2(f-v)u +  2-v-f)dvdu\nonumber\\
f&\equiv&f(v,u)= 1-\frac{v}{2} -\sqrt{ (1-\frac{v}{2})^2-\frac{1-v}
{1-u} } \nonumber\\
 u&=& \bar{n}_1\cdot \bar{n}_2 \ \ \ \bar{n}_i =
\frac{\bar{k}_i}{\omega_i} \
\ \ umax=\frac{(\frac{v}{2})^2}{(1-\frac{v}{2})^2}\nonumber\\  v&=&
\frac{\omega_3}{m_e}\ \ x= \frac{m_{\pi_L}}{2m_e}
 \end{eqnarray}

\noindent $\omega_i$ and $\bar{k}_i$ denote the energies of photons,
$u$ denotes the cosine of the angle between first and second photon and $v$
is the energy of the third photon using electron mass as unit.
The condition
$\Delta \Gamma/\Gamma =10^{-3}$ gives for the parameter $f_{\pi_L}$
the value $f_{\pi_L}\simeq 7.9  \ keV$.
The existence of at least 3 states identifiable as
$\sigma$ scalars,  suggests the existence of several leptopion states
in one-one correspondence with sigma scalars (string
model satellites). Since  these states contribute to decay anomaly
 additively
the estimate for $f_{\pi_L}$ assumed to scale as $ m_{\pi_L}$  increases and
one
obtains $f_{\pi_L}\simeq 11 \ keV$ for the lightest leptopion state. From
the PCAC
relation one obtains for $sin(\theta_e)$ the upper bound
$sin(\theta_e)\leq x \cdot
10^{-4}$ assuming $m_{ex}\geq 1.3 \ MeV$ (so that $e_{ex}\bar{e}$ decay is
 not
possible), where $x=1.2$ for single leptopion state and $x=1.4$ for 3
leptopion states.

 \vm

Leptopion photon interaction  implies also a new contribution to photon-photon
scattering. Just at the threshold $E=m_{\pi_L}/2$ the creation of leptopion in
photon photon scattering is possible  and the appearence of leptopion as
virtual particle gives resonance type behaviour to photon photon scattering
near $s=m_{\pi_L}^2$.  The total photon-photon cross section in zero decay
width approximation is given by

\begin{eqnarray}
\sigma &=& \frac{\alpha^4}{2^{14}(2\pi)^6 } \frac{E^6}{
f^4_{\pi_L}(E^2-\frac{m_{\pi_L}^2}{4})^2}
\end{eqnarray}

\noindent  The last column of the table 1 gives the value of
the cross section at resonance.

\vl

\begin{tabular}{||c |c | c|c |c|c||}\hline \cline{1-6}
$N$& $Op/10^{-3}$    &$f_{\pi_L}/keV$
  &$ sin(\theta_e)(m_{ex}/1.3 \ MeV)^{1/2} $& $\Gamma (\pi_L)/keV$
 &$\sigma(\gamma\gamma)/\mu b$ \\ \hline
1& 1   & $ 7.9 $   &$ 1.2\cdot 10^{-4}$&$ .51   $    &$ .03  $ \\  \hline
3 & 1  & $ 11.0$   &$ 1.4\cdot 10^{-4}  $ &$ .27   $ &$ .007 $ \\ \hline
3 & 5  & $ 4.9$    &$ 3.1\cdot 10^{-4} $ &$ 1.3   $ &$ .18  $  \\ \hline
\hline
\end{tabular}
\vl

Table 1:  The dependence of various quantities on the number of leptopion type
states and Op anomaly. $N$ refers to the number of leptopion states and $Op$
denotes leptopion anomaly. Last column gives the value of the photon-photon
scattering cross section at resonance.

\subsection{Spontaneous vacuum expectation of leptopion field
as source of  leptopions}

The basic  assumption  in   the model of  leptopion and leptohadron
 production
is the spontaneous generation of leptopion vacuum expectation value
in strong nonorthogonal electric and magnetic fields. This assumption
is in fact very natural in TGD.

\vm

a) The well known relation \cite{Iztyk} expressing pion field as a sum
 of
the divergence of axial vector current and anomaly term generalizes to
the case of leptopion

\begin{eqnarray}
\pi_L&=&\frac{1}{f_{\pi_L} m^2(\pi_L)}
(\nabla\cdot j^{A} +\frac{\alpha_{em}}{2\pi}E \cdot B)
\end{eqnarray}

\noindent  In the case of leptopion case
the value of $f_{\pi_L}$ has been already deduced from PCAC argument.
 Anomaly term gives rise to pion decay to two photons  so that one
obtains an estimate for the lifetime of the leptopion.

\vm

This relation is taken as the basis for the model describing also
the production of leptopion in external electromagnetic field.
The idea is that the presence of external electromagnetic field
gives rise to a vacuum expectation value of leptopion field. Vacuum
expectation is obtained by assuming that the vacuum expectation
value of axial vector current vanishes.

\begin{eqnarray}
\langle vac\mid\pi\mid vac\rangle&=&KE \cdot B
\nonumber\\
K&=&\frac{\alpha_{em}}{2\pi f(\pi_L)m^2(\pi_L)}
\end{eqnarray}

\noindent Some comments concerning this hypothesis are in order here:\\
i) The basic
hypothesis making possible  to avoid large parity breaking effects in atomic
and molecular physics is that p-adic condensation levels with length scale
$L(n)<10^{-6} \ m$ are purely electromagnetic in the sense that nuclei feed
their $Z^0$ charges on condensate levels with $L(n)\geq 10^{-6} \ m$.
The absence of  $Z^0$ charges  does not however exclude the possibility of
the classical  $Z^0$ fields induced by the nonorthogonality  of the
ordinary electric and
magnetic fields (if $Z^0$ fields vanish $E$ and $B$ are orthogonal in TGD
(cite{TGD}).  \\ ii) The nonvanishing vacuum expectation value of the leptopion
field  implies parity breaking in atomic length scales. This is understandable
from basic principles of TGD  since classical
$Z^0$ field has parity breaking axial coupling to electrons and protons. The
nonvanishing classical leptopion field is in fact more or less equivalent with
the  presence of classical $Z^0$ field.

\vm

b)  The amplitude for the production of leptopion with four momentum
$p=(p_0,\bar p)$ in an  external electromagnetic field
can be deduced by writing leptopion field as sum of classical and quantum
parts: $\pi_L = \pi_L(class)+ \pi_L(quant)$ and
by decomposing    the  mass term into interaction term plus
c-number term and standard mass term:

\begin{eqnarray}
\frac{m^2(\pi_L) \pi_L^2}{2}&=&  L_{int}+ L_0\nonumber\\
L_0&=& \frac{m^2(\pi_L)}{2}(\pi^2_L(class)+  \pi^2_L(quant))\nonumber\\
L_{int}&=& m^2(\pi_L)\pi_L(class)\pi_L(quant)
\end{eqnarray}

\noindent Interaction Lagrangian corresponds to  $L_{int}$
linear in leptopion oscillator operators.  Using standard
LSZ reduction formula and normalization conventions of \cite{Iztyk}   one
obtains  for the probability amplitude for creating  leptopion of momentum
$p$
 from vacuum the expression

\begin{eqnarray}
A(p)&\equiv & \langle  a (p) \pi_L\rangle =(2\pi)^3 m^2(\pi_L) \int f_p (x)
 \langle vac\mid\pi\mid vac\rangle d^{4}x \nonumber\\
f_p&=&e^{ip\cdot x}
\end{eqnarray}

\noindent   The  probability for the production of leptopion in phase space
volume element $d3p$ is obtained by multiplying with the density of
states factor $d^3n = \frac{d^3p}{(2\pi)^3 2E} $:

\begin{eqnarray}
dP&=&A\mid U \mid^{2}\frac{ d^{3}p}{2E_p}
\nonumber\\
A&=& (\frac{\alpha_{em}}{2\pi f(\pi_L)})^{2} \nonumber\\
U&=& \int e^{ip\cdot x}E \cdot B d^{4}x
\end{eqnarray}

\noindent The first  conclusion  that one can draw is
that nonstatic electromagnetic fields are required for leptopion creation
since in static fields energy conservation forces leptopion to have
zero energy and thus prohibits real  leptopion production.
In particular, the spontaneous creation leptopion in static Coulombic  and
 magnetic  dipole fields
of nucleus is impossible.

\subsection{Sigma model and creation of leptohadrons in  electromagnetic
fields}

\subsubsection{ Why sigma model approach?}

For several reasons it is necessary to generalize
the model for leptopion production  to a   model for  leptohadron
production.\\ a) Leptopions probably correspond to resonances with mass
$m(\pi_L) \simeq 1.062 \ MeV $ \cite{Chodos}  and decay mostly to photon photon
pairs so that  $1.8 \ MeV$ resonance should correspond to some other
leptomeson.
Besides pseudoscalars $\eta_L$ and $\eta'_L$ one can consider vector bosons
$\rho_L$ and $\omega_L$ and  scalar particle and its radial excitations
 $\sigma_L$ as candidates for  the observed  resonances.   \\
b) A model for the production of leptohadrons is obtained from an effective
action describing the strong and electromagnetic interactions between
leptohadrons.   The simplest model is sigma model describing the interaction
between leptonucleons, leptopion and a  hypothetical scalar particle
$\sigma_L$ \cite{Iztyk}.  This model realizes leptopion
field as a divergence of the axial current and gives the standard relation
between $f_{\pi_L},g$ and $m_{ex}$.
 All couplings of the model are related to the masses of
$e_{ex},\pi_L$ and $\sigma_L$.
The generation of leptopion vacuum expectation value in the proposed manner
takes place via triangle anomaly diagrams in the external electromagnetic
field.\\ c) If needed the model can be
generalized to contain terms describing also other leptohadrons.  The
generalized model should contain also vector bosons $\rho_L$ and $\omega_L$
as well as pseudoscalars $\eta_L$ and $\eta'_L $
and radial excitations of $\pi_L$ and $\sigma_L$.  An  open question is
whether
  also $\eta$
 and $\eta'$ generate
vacuum expectation value proportional to $E\cdot B$. \\
d)  The following argument suggests  that the most plausible
identification of $1.8 \ MeV$ resonance is as $\sigma_L$ so that sigma
model indeed provides a satisfactory description of the situation. \\
i) The mass of $e_{ex}$ must be so large that the  decays  to
$e_{ex}\bar{e_{ex}}$ pairs are forbidden (they would lead to nonsensically
large decay width). The
most plausible production mechanism for $e^+e^-$ pairs is the  decay of
leptomeson  to $e^+e^-$ pair but one cannot exclude the decay to
$e_{ex}\bar{e}$ and  subsequent decay
 $e_{ex}\rightarrow e+\gamma$.\\
ii)
Ortopositronium decay width gives $f_{\pi_L}\sim .0021 \ MeV$ and from this
one can deduce  an upper bound for
leptopion production cross section in an external electromagnetic field.
The calculation of leptopion production cross section shows that leptopion
production cross section is
 somewhat smaller than the observed $e^+e^-$ production
cross section,  even  when one tunes the values of the various parameters.
Since pseudoscalars are expected to decay mostly to photon pairs leptopions,
$\eta_L$ and $\eta'_L$ as main  source of $e^+e^-$ pairs are unprobable.\\
iii)   The direct production of the  pairs  via the  interaction term\\
$gsin(\theta_e)\bar{e}_L\gamma_5e_{ex}\pi_L(cl)$ from is much slower process
than the production via the meson decays  and does not give rise to resonant
structures since
mass squared spectrum  for pairs forms continuum.  Also  the production via
the  $\bar{e}e_{ex}$ decay of virtual leptopion  created from classical field
 is slow process since it involves $sin^2(\theta_e)$. \\
iv)   $e^+e^-$ production  can proceed also via the
creation of many particle states. The simplest candidates are $V_L+\pi_L$
states created via $\partial_{\alpha}\pi_LV^{\alpha}\pi_L(class)$ term in
action and  $\sigma_L +\pi_L$ states created via the the
$k\sigma_L\pi_L\pi_L(class)$ term in the sigma model action.
  The
production cross section via the decays of vector mesons  is certainly very
small since the production vertex involves the inner product of vector boson
3 momentum with its polarization vector and the situation is nonrelativistic.
\\
 v) The pleasant surprise is
that the production rate  for $\sigma_L$  meson is large since the coupling
$k$ turns out to be given by $k= (m^2_{\sigma_L}-m^2_{\pi_L})/2f_{\pi_L}
$ and
is  anomalously large for the
 value of $f_{\pi_L}\geq .0079 \ MeV$   derived from ortopositronium anomaly:
 $k \sim 336m(\pi_L)$ for $f_{\pi_L}\sim 7.9 \ keV$.
The resulting additional factor
in  the production cross section compensates the reduction
factor
coming from two-particle phase space volume and gives a cross section, which
is rather  near to  the maximum value of the observed cross section.

\subsubsection{Simplest sigma model}

A detailed description of the sigma model can be found in \cite{Iztyk} and it
suffices to outline  only  the crucial features here.\\
a) The action of leptohadronic sigma model reads as

\begin{eqnarray}
L&=& L_S+ c\sigma_L\nonumber\\
L_S&=& \bar{\psi_L} (i\gamma^k\partial_k +g(\sigma_L +i\pi_L\cdot \tau
\gamma_5))\psi_L +\frac{1}{2}((\partial \pi_L)^2+ (\partial \sigma_L)^2)
\nonumber\\
&-&\frac{\mu^2}{2}(\sigma_L^2+ \pi_L^2)
-\frac{\lambda}{4}(\sigma_L^2+\pi_L^2)^2 \end{eqnarray}

\noindent $\pi_L$ is isospin triplet and $\sigma_L$ isospin singlet. $\psi_L$
is isospin doublet with electroweak quantum numbers of electron and neutrino
($e_{ex}$ and $\nu_{ex}$).  The model
allows $so(4)$ symmetry. Vector current is conserved but for $c\neq 0$ axial
current generates divergence, which is proportional to pion field:
$\partial^{\alpha}A_{\alpha}= -c\pi_L$.\\
 b) The presence of the  linear term implies that $\sigma_L $ field
generates vacuum expectation value $\langle 0\vert\sigma_L \vert 0\rangle=v$.
When the action is written in terms of new quantum field $\sigma'_L=
\sigma_L-v$ one has

\begin{eqnarray}
L&=& \bar{\psi}_L (i\gamma^k\partial_k + m + g(\sigma'_L +i\pi_L\cdot \tau
\gamma_5))\psi_L +\frac{1}{2}( (\partial \pi_L)^2+ (\partial \sigma_L')^2)
\nonumber\\ &-&\frac{1}{2}  m^2_{\sigma_L}(\sigma'_L)^2 - \frac{m^2_{\pi_L}}{2}
\pi_L^2\nonumber\\ &-&\lambda v\sigma'_L((\sigma'_L)^2+\pi_L^2)
-\frac{\lambda}{4}((\sigma_L')^2+\pi_L^2)^2  \nonumber\\
\
\end{eqnarray}

\noindent  The  masses are given by

\begin{eqnarray}
m^2_{\pi_L}&=& \mu^2 +\lambda v^2\nonumber\\
m^2_{\sigma_L}&=& \mu^2 +3\lambda v^2\nonumber\\
m&=& -gv
\end{eqnarray}

\noindent These formulas relate the parameters $\mu,v,g$ to leptohadrons
masses.\\ c) The requirement that $\sigma'_L$ has vanishing vacuum
expectation implies in Born approximation

\begin{eqnarray}
c-\mu^2v-\lambda v^3&=&0
\end{eqnarray}
\noindent which implies

\begin{eqnarray}
f_{\pi_L}&=& -v= -\frac{c}{m^2(\pi_L)}\nonumber\\
m_{ex}&=& gf_{\pi_L}
\end{eqnarray}

\noindent Note that $e_{ex}$ and $\nu_{ex}$ are predicted to have identical
masses in this approximation. \\
d) A new feature  is the generation of the leptopion vacuum expectation value
in an  external electromagnetic field (of course,  this is possible for  the
ordinary pion field, too!).   The
vacuum expectation is generated via the triangle anomaly diagram in a manner
identical to the generation of a nonvanishing photon-photon decay amplitude
and is proportional to the instanton density of the electromagnetic field. By
redefining the pion field as a  sum
$\pi_L =\pi_L(cl)+\pi'_L$ one obtains effective action describing the
creation of the leptohadrons in strong electromagnetic fields.\\
 e) As
far as the production of $\sigma_L\pi_L$ pairs is considered,  the
interaction term  $\lambda v\sigma'_L\pi_L^2 $ is especially interesting
since it leads to the creation of $\sigma_L\pi_L$ pairs via the interaction
term $k\lambda v\sigma_L'\pi_L(cl)$.
 The coefficient of this term can be
expressed in terms of the  leptomeson masses and $f_{\pi_L}$:

\begin{eqnarray}
k&\equiv& 2\lambda v= \frac{m^2_{\sigma_L} -m^2_{\pi_L}}{2f_{\pi_L}}
\end{eqnarray}

\noindent The large
value of the coupling  ($k\sim 336m_{\pi_L}$
for $f_{\pi_L}=7.9 \ keV$)  compensates the reduction of the
production rate coming from the smallness of two-particle phase space volume
as compared with single particle-phase space volume.

\subsubsection{ How to generalize the sigma model approach?}

The simplest sigma  model containing only
pion and $\sigma$ particle is certainly an  overidealization
since three  resonances
 at energies $1.63, 1.77 $ and $1.83 \ MeV$ rather than just  one
have been identified (besides
leptopion at $1.062 \ MeV$).  This suggests a generalization of  the
simplest sigma model approach.\\
a)  The production of $\sigma$ particle together with some other
particles is necessary in order to obtain large enough
$e^+e^-$  production
cross section without ad hoc assumptions about the values of coupling
constants.\\ b) The first,rather unprobable,  possibility
 is  that some other pseudoscalars besides $\pi_L$
 can be produced in association with $\sigma_L$ and the decay of
these states gives rise to $e_{ex}\bar{e}$ pairs since  the direct
decay to $e^+e^-$ is too slow as compared with $\gamma\gamma$ decay.
This requires that the mass of  $e_{ex}$ is below $1.12 \ MeV$.
  The pseudoscalars are probably not
 the leptonic
counterparts of $K_0$, $\eta$ $\eta'$ meson: these pseudoscalars
 contain $g=1$ color octet leptons  (counterparts of strange
quarks) and their masses are expected to be larger than the observed
masses.
\\ c)  The second possibility is that the
states with mass above $1.6 \ MeV$ correspond to radial excitations of
leptosigma.  In  string model  radial excitations correspond to
satellite trajectories of the
 highest  Regge
trajectory and the states obey the mass formula

\begin{eqnarray}
M^2(J,n)&=&M_0^2 + T(J-2n)
\end{eqnarray}

\noindent Here $J=0$ is the spin of the resonance,  the integer $n$
labels  the satellite in question and $T$ denotes leptohadronic string
tension of order
 one MeV. $2n$ appears instead of $n$ in the formula to guarantee
that states are scalars.  The number of satellites is clearly finite.
The  masses of resonances above $1.6 \ MeV$ are in a satisfactory
approximation evenly spaced, which suggests that the condition
$T<<M_0^2$ holds true. This scenario leads to the prediction of
altogether $9$ satellites of $\sigma_0$ with mass $\sim 1.83 \ MeV$.
The  production rate for satellite is  proportional to the factor
$(m^2_{\sigma_L(n)}-m^2_{\pi_L})^2$ so that production probability
for the  lowest mass satellites is smaller and might explain why these
states have not been observed.\\ d) Same formula predicts satellites
for leptopion, too.
These states  have masses below $2m_e$ and
can decay to two-photon states only. Stability of electron against
$\pi_L+\nu$ decay implies $m_{\pi_L}\geq m_e$ and the smallness
of Op anomaly as well as the experimental absence of the decay
$Op\rightarrow \pi_L+\gamma$ implies $m_{\pi_l}>2m_e$ so that
 $1.062 \ MeV$
state must be the lightest leptopion state.  If leptopion and sigma
satellites
form
$so(4)$ multiplets  this means that  $\pi_L$ has  at least
2 satellites. The estimate for  the masses
of leptopion and sigma states are given in table below for string tension
$T=0.178 \ MeV$.

\vm
\begin{tabular}{||c |c | c|c ||}\hline \cline{1-4}
$n$& 0    & 1       &2 \\ \hline
$m(\pi_L)/MeV$& 1.36  &1.22  & 1.062    \\  \hline
$m(\sigma_L)/MeV$ &1.83  &1.73   & 1.62  \\ \hline \hline
\end{tabular}

\vl

Table 2. String model mass estimate for the satellites of $\pi_L$ and
$\sigma_L$.

\vm
The naivest generalization of  the sigma model means  the
 arrangement $\sigma_L-\pi_L$ pairs associated with
various satellites to $so(4)$ multiplets so that each pair gives its
own
 contribution to the sigma model  action. The simplest assumption is
that the sigma model  couplings $g$ and $\lambda$ associated  with
various satellites are identical and $c$ scales as $m_{\pi_L}^3$.
 It seems natural to associate to given a given  meson multiplet the
corresponding leptonucleon satellite: the mass of the leptonucleon
would scale as $m_{\pi_L}$ in the simplest scenario.

\subsection{ Classical model  for  leptopion production }

The nice feature of the model (and its possible generalizations)
is that the production amplitudes associated with all leptohadron production
reactions  in external electromagnetic field are
proportional to the leptopion production amplitude and apart from phase space
volume factors
  production cross sections
are expected to be   given by
leptopion production cross section.
Therefore it makes sense to construct  a detailed model for leptopion
production despite the fact that leptopion decays probably contribute only a
very small fraction to the observed $e^+e^-$ pairs.

\vm

Angular momentum barrier makes the production of leptomesons  with
orbital angular momentum $L>0$  unprobable. Therefore the observed
resonances are expected to be $L=0$ pseudoscalar states. Leptopion
production has two signatures which any realistic model should
reproduce.\\
a) Data are consistent with the assumption that states are produced
at rest in cm frame.\\
b) The production probability has a peak in a narrow region of velocities
of colling nucleus around the velocity needed to overcome Coulomb
barrier in head on collision. The relative width of the velocity
peak is of order $\Delta \beta/\beta\simeq\cdot 10^{-2}$ \cite{Cowan}.
In Th-Th system
\cite{Cowan} two peaks at projectile energies 5.70 MeV and 5.75 MeV per
nucleon have been observed.  This suggests that some kind of diffraction
mechanism based on the finite size of nuclei is at work.\\
In this section  a model treating nuclei as point like charges and
nucleus-nucleus collision purely classically is developed. This
model yields  qualitative predictions in agreement with the signature a)
but fails to reproduce the possible diffraction behaviour although
one can develop argument for understanding the behaviour above Coulomb wall.

\vm

The general expression for the amplitude for creation of leptopion in
external electric and magnetic fields has been already derived.
Let us now specialize to the case of heavy ion collision. We consider
the situation,  where the scattering angle of the colliding nucleus
is measured. Treating  the collision completely classically we can
assume that collision occurs with a well defined value of  the impact
parameter in a fixed scattering plane.
The coordinates are chosen so that target nucleus is at rest at the
origin of the coordinates and colliding nucleus moves in z-direction
in y=0 plane with velocity $\beta$. The scattering angle of the
scattered nucleus is denoted by $\alpha$, the velocity of the lepto-
pion by v and the direction angles of leptopion velocity by
$(\theta,\phi)$.

\vm

The minimum value of the impact parameter for the Coulomb collision
of point like charges is given by the expression

\begin{eqnarray}
b&=& \frac{b_{0} cot(\alpha/2)}{2}
\nonumber\\
b_{0}&=&\frac{2Z_1Z_2\alpha_{em}}{M_R\beta^2}
\end{eqnarray}

\noindent where $b_{0}$ is the expression for the distance of the closest
approach in head on collision. $M_R$
denotes the reduced mass of the nucleus-nucleus system.

\vm

To estimate the amplitude for leptopion production the following
simplifying assumptions are made. \\
a) Nuclei can be treated as point like charges.
This assumption is well motivated,  when the impact parameter of the
collision is  larger than the critical impact parameter given by the
sum of radii of the colliding nuclei:

\begin{eqnarray}
b_{cr}&=&R_1+R_2
\end{eqnarray}

\noindent For scattering angles that are sufficiently large the values of the
impact parameter do not satisfy the above condition  in the region
of the velocity peak.  p-Adic considerations lead to the conclusion
that nuclear condensation level corresponds to prime $p\sim 2^{k}$ , $k=113$
($k$ is prime). This  suggest that nuclear radius
 should
be replaced by the size $L(113)$ of the p-adic convergence cube  associated
with nucleus \cite{padTGD}: $L(113)\sim 2.26\cdot 10^{-14} \ m$ implies that
cufott radius
is $b_{cr}\sim 2L(113)\sim 5.2 \cdot10^{-14} \ m$.
\\
b) Since the velocities are nonrelativistic (about $ 0.12  c$) one can
treat the motion of the nuclei nonrelativistically and  the nonretarded
electromagnetic fields associated with the exactly known classical
orbits can be used. The use of classical orbit doesn't take into
account recoil effect caused by leptopion production. Since the mass
ratio of leptopion and the reduced mass of heavy nucleus system is
of order $10^{-5}$ the recoil effect is however negligible.\\
c) The model simplifies considerably, when the orbit is idealized with
a straight line with impact parameter determined from the condition
expressing scattering angle in terms of the impact parameter. This
approximation is certainly well founded for large values of impact
parameter. For small values of impact parameter the situation is quite
different and an interesting problem is  whether the contributions of
long range radiation  fields created by accelerating nuclei in
 head-on collision could give large contribution to leptopion production
rate. On the line connecting the nuclei the electric part of the
radiation field created by first nucleus is indeed parallel to the magnetic
part of the radiation field created by second nucleus.
In this approximation  the instanton density in the rest frame of
the target nucleus is just the scalar product of the Coulombic
electric field E of the target nucleus and of the magnetic field B of
the colliding nucleus obtained by boosting it from the Coulomb field
of nucleus at rest.

\subsubsection{Cutoff length scales in the  classical model }

The differential cross section in the  classical model can be written as

\begin{eqnarray}
dP&=&
K_0\mid U(b) \mid^{2}\frac{ d^{3}p}{2(2\pi)^3E_p} 2\pi bdb
\nonumber\\
K_0&=& (\frac{\alpha_{em}}{2\pi f(\pi_L)})^{2} \nonumber\\
U(b,p)&=& \int e^{ip\cdot x}E \cdot B d^{4}x
\end{eqnarray}

\noindent where $b$ denotes impact parameter.
In the calculation of the total cross section one must introduce some cutoff
radii.

\vm

Consider first the choice of the lower cutoff length scale $b_{cr}$.\\
a) Since leptopion production has maximum at
energy near Coulomb wall suggest that the finite size of the colliding
nuclei
 might play  important role in the
collision for the values of the scattering angle and velocity
considered. \\
b) Lower impact parameter cutoff makes sense if
the contribution of small impact parameter collisions to the
production amplitude is small. This seems to be the case. For head on
collision $E\cdot B$  vanishes identically and
 by continuity  leptopion production
 amplitude must decrease
with increasing value of scattering angle for small values
of  impact parameter.   The value
of $b_{cr}$  should
lie somehere between $2\cdot 10^{-14} \ m $ and $10^{-13} \ m$  but
 its exact value  is subject to considerable uncertainty.
For impact parameters below the value of two nuclear radii point like nature
of nuclei is not a good approximation and the value of
$E\cdot B$ becomes more or less random
in the interaction region and  Fourier transform of $E\cdot B$
becomes small. For fixed scattering angle of nuclei
this could explain why production rate becomes small above Coulomb wall.
What happens is that for fixed value of $\theta$ the impact parameter $b$
becomes smaller than $b_{cr}Ê\sim 2L(113)$, when critical value of collision
velocity
$\beta$ is reached
and $E \cdot B$ becomes random in the interaction region.
\\
c)   Alternatively,   lower cufoff length scale
could result from the requirement that maximum scattering angle is so  small
for the approximation of  linear nuclear motion to make sense. Assume
that the  maximum scattering angle  is $\theta (max) = n\theta (min)$,
 $\theta (min)=
2Z_1Z_2\alpha/M_Ra\beta^2 \sim 4 \cdot 10^{-2}/A$ for $a\sim 10^{-10} \ m$
with   $\theta (max) \sim .1$. This gives
 $b_{cr}\sim  10^{-13} \ m$. This scale is by a factor of order
two larger than that lower cutoff length scale given by the p-adic argument
so that there seems to be a region of impact parameters,  where the
approximation of linear motion need not be good.  If the contribution of the
large angle collisions having
$\theta\geq n\theta_1$ to the production amplitude is small then
the decrease of the production probability could occur already below the
Coulomb wall.

\vm

   Consider next the constraints on  the upper cutoff length scale.\\
a) The production  amplitude turns out to
decrease exponentially  as a function of impact parameter $b$
unless leptopion is produced in scattering plane. The contribution of
leptopions produced in scattering plane however gives divergent contribution
to the  total cross section integrated over all impact parameter values
 and upper cutoff length scale $a$ is necessary. If
one
considers scattering with scattering angle between specified limits this is
 of course  not
a problem of classical model.
 \\
b) Upper cutoff length scale  $a$ should be certainly smaller than the
interatomic distance. A more stringent upper bound for
$a$  is  the size $r$
 of atom defined as the distance  above  which atom looks essentially neutral:
a rough extrapolation from hydrogen atom
 gives $r\sim a_0/Z^{1/3}\sim 1.5\cdot 10^{-11} \ m$ ($a_0$ is Bohr radius
of hydrogen atom).
Therefore
cutoff scale is between  Bohr radius $a_{0}/Z \sim .5\cdot 10^{-12} \ m$ and
$r$. \\ c)   One could perhaps understand the appearence of the  upper cutoff
 length scale
of order $10^{-11}\ m$ from p-adic considerations.
Leptopions have primary p-adic
condensation level $k=127$ and are condensed on level $k=131$. Leptopions
are created at condensate level $k=131$ in
the classical electromagnetic fields of the colliding nuclei.
$L(131)$ serves as a natural infrared cutoff for p-adic physics
at leptopion  condensation level $k=131$ so that one must conclude
that leptopion production rate should be calculated from p-adic physics.
One can however hope that the model based on real numbers gives satisfactory
description of the situation, when the presence of the  p-adic cutoff length
scale
is taken into account.  Notice that
  $a=10^{-11} \ m $   corresponds also  to de Broglie wavelength for
the  leptopions of velocity $v_{cm}\sim .1$.

\subsubsection{Production amplitude}

The  Fourier transform of $E\cdot B$  can be expressed
as a convolution of Fourier transforms of E and B
and the resulting expression for the amplitude reduces by residue calculus
(see APPENDIX)
 to the following
general form

\begin{eqnarray}
U&=&N(CUT_1+CUT_2)\nonumber\\
N&=&\frac{i}{ (2\pi)^7}
\end{eqnarray}

\noindent  where nuclear  charges are such that Coulomb potential is $1/r$.
The contribution of the first cut for  $\phi \in [0,\pi/2]$
 is given by the expression

\begin{eqnarray}
CUT_1&=& \frac{1}{2}sin(\theta)sin(\phi)\int_0^{\pi/2}
 exp(-\frac{cos(\psi)x} {sin(\phi_0) }) A_1d\psi
\nonumber\\
A_1&=& \frac{Y_1}{X_1}\nonumber\\
Y_1&=& sin(\theta)cos(\phi)+
iKcos(\psi)\nonumber\\
X_1&=& sin^2(\theta)(sin^2(\phi)-cos^2(\psi)) +K^{2}-2iKsin(\theta)
cos(\psi)cos(\phi)\nonumber\\
K&=&\beta\gamma (1- \frac{v_{cm}cos(\theta)}{\beta})
\nonumber\\
sin(\phi_0)&=& \frac{\beta\gamma}{am(\pi_L)\gamma_1},\ \   \
\gamma_1= \frac{1}{\sqrt{1-v^2}},  \ \   \ \gamma = \frac{1}
{\sqrt{1-\beta^2}},
 \  \   \ v_{cm}=\frac{2v}{(1+v^{2})} \nonumber\\
 x&=&\frac{b}{a}
\end{eqnarray}

\noindent  The dimensionless  variable
$x=b/a$ is the ratio of the  impact parameter to the upper
cutoff radius $a$.

\vm

The contribution of the second cut is given
 by the expression

\begin{eqnarray}
CUT_2&=& \frac{1}{2}usin(\theta)sin(\phi)
exp(ir_1sin(\theta)cos(\phi)x)\int_0^{\pi/2}
exp(-r_2cos(\psi)x)A_2d\psi \nonumber\\
A_2&=& \frac{Y_2}{X_2}\nonumber\\
Y_2&=& sin(\theta)cos(\phi) u - 2icos(\psi)(\frac{w}{v_{cm}}+
\frac{v}{\beta}sin^2(\theta)
cos(2\phi))\nonumber\\
X_2&=&sin^2(\theta)(\frac{sin^2(\phi)}{\gamma^2} -u^{2}cos^2(\psi)
+ \beta^2 (v^{2}sin^2(\theta)-\frac{2vw}{v_{cm}}cos^2(\phi))) \nonumber\\
&+& \frac{w^2}{v_{cm}^2}
+2iu\beta  sin(\theta)cos(\phi)(vsin^2(\theta)cos(\phi)-\frac{wcos(\psi)
}{v_{cm}}) \nonumber\\
u&=&1-\beta vcos(\theta)\  \ \ w = 1-rcos(\theta)\nonumber\\
Êr_1&=& \frac{\beta  v\gamma}{sin(\phi_0)}\ \  \ r_2= \frac{\gamma}
{sin(\phi_0)}  \ \  \ r= \frac{v_{cm}}{\beta} \nonumber\\
 x&=& \frac{b}{a}
\
\end{eqnarray}

\noindent The denominator $X_2$ has no poles  in the  physical
region and the contribution of the
second cut is therefore finite.  Besides this the exponential damping
makes the integrand small everywhere expect in the vicinity of $cos(\Psi)=0$
and for small values of the impact parameter.

\vm

 Using the symmetries

\begin{eqnarray}
A(p_x ,-p_y )&=&-A(p_x ,p_y )
\nonumber\\A(-p_x ,-p_y )&=&\bar{A}(p_x ,p_y )
\end{eqnarray}

\noindent of the amplitude one can calculate the amplitude for other values
of $\phi$.

\vm

$CUT_1$ gives the singular contribution to the amplitude.
The reason is that  the factor $X_1$ appearing in denominator of cut
term vanishes,  when the conditions

\begin{eqnarray}
cos(\theta)&=& \frac{\beta}{v_{cm}}\nonumber\\
sin(\phi)&=&cos(\psi)
\end{eqnarray}

\noindent are satisfied.  In forward direction this condition tells that
 z- component of the
leptopion momentum in velocity center of mass coordinate system
vanishes.  In laboratory this condition means that the leptopion
moves in certain cone defined by the value of its velocity. The
condition is possible to satisfy only above the threshold
$v_{cm}\geq \beta$.

\vm

\noindent For $K=0$ the integral reduces to the form

\begin{eqnarray}
CUT_1 &=& \frac{1}{2}cos(\phi)sin(\phi) \lim_{\varepsilon\rightarrow 0}
\frac{\int_0^{\pi/2} exp(-\frac{cos(\psi)}{sin(\phi_0)} ) d\psi}
{(sin^2(\phi) -
cos^{2}\psi +i\varepsilon )}\nonumber\\
\
\end{eqnarray}

\noindent One can estimate the singular part of the integral by replacing the
exponent term with its value at the pole. The integral contains two parts:
the first part is principal value integral and second part can be regarded
as integral over a small semicircle going around the pole of integrand in
upper half plane.
The remaining integrations
can be performed using elementary calculus
 and one obtains for the singular cut contribution
the approximate expression

\begin{eqnarray}
CUT_1&\simeq& e^{-(b/a) (sin(\phi)/sin(\phi_0))} (
\frac{ln(X)}{2} + Ê\frac{i\pi}{2} \nonumber\\
X&=&\frac{((1+s)^{1/2}+(1-s)^{1/2})}{((1+s)^{1/2}-(1-s)^{1/2})}\nonumber\\
s&=&sin(\phi)\nonumber\\
sin(\phi_0)&=& \frac{\beta \gamma}{\gamma_1 m(\pi_L)a}
\end{eqnarray}

\noindent The principal value contribution to the
amplitude diverges logarithmically for $\phi=0$ and dominates over 'pole'
contribution for small values of $\phi$.   For finite  values
of impact parameter the amplitude decreases expontially as a function
of $\phi$.

\vm

If the singular term  appearing in $CUT_1$Ê
indeed gives the dominant contribution to the
leptopion production one can make  some conclusions concerning
the properties of the production amplitude.   For given leptopion cm
velocity $v_{cm}$
the production  associated with the singular peak
is predicted to occur mainly in the cone $cos(\theta)=\beta/v_{cm}$:
 in
forward direction this corresponds to the  vanishing of the
z-component of the leptopion momentum in velocity center of mass
frame. Since the values of $sin(\theta)$ are of order
$.1$ the transversal momentum is small and
production occurs almost at rest in cm frame as observed.
In addition, the singular
 production  cross  section is concentrated
in the production plane ( $\phi=0$)  due to the  exponential
dependence of the singular  production
amplitude on the   angle
$\phi$  and impact parameter
and the presence of the logarithmic singularity.
 The observed leptopion velocities are in the range  $ \Delta v/v\simeq 0.2$
\cite{Cowan}
and this corresponds to the angular width $\Delta\theta\simeq 34$
degrees.

\vm

 These conclusions are justified by the numerical calculation of leptopion
production probability $P(b)$
described in the Appendix. In the figure \ref{clacross}  the leptopion
 differential cross section  $\frac{d\sigma}{d\Omega dv} = 2\pi\int P(b)bdb$
integrated over impact parameters in the range $(b_{cr},a)$
in U-U collision is plotted as
function of scattering of direction  angles $(\theta,\phi)$ of leptopion
momentum.
 The values of various parameters are
 $Z_1=Z_2=92$, $a=10^{-11} \
m$, $ b_{cr}=4\cdot 10^{-14} \ m$,
$(\beta,v) =(.102 ,.106 )$ ($v$ is cm velocity),
$(\theta_0, \phi_0)=(16.0 ,.002 ) \ degrees$. The upper cutoff has values
$a=10^{-12} \ m$ and $a=10^{-11} \ m$.
  Figures  a) and  c) give overall view of the differential cross section
and figures b) and d) display  the
behaviour of the differential cross section in the singular region.  For fixed
value of $v$ the cross
section is peaked to momenta in scattering plane near $\theta=\theta_0$.

\begin{figure}[htb]
\leavevmode
\centering
\vspace*{1cm}
\label{clacross}
\caption{Dependence of the leptopion  differential
production cross section $\frac{d\sigma}{d\Omega dv}$  on  angles $\theta$
and $\phi$ in the classical model.
a) Total view for $a=10^{-12} \ m$. b) Singular region for $a=10^{-12}
 \ m$.
c) Total view for $a=10^{-11} \ m$. d) Singular region for $a=10^{-11} \ m$.
Various parameter values are given in the text. }
 \end{figure}

\clearpage
\subsubsection{Leptopion production cross section in the  classical model}

There are no free parameters in the model and the  comparison
of the predicted leptopion  production  cross section with
the  measured $e^+e^-$ production cross section serves as a stringent   test
of  the theory.
The largest experimental value  for the
production cross section $\sigma_{exp}  $ is
$\sigma_{exp}(e^+e^-) \sim 5 \cdot \mu b$
\cite{Tsertos}.
The differential production cross section is concentrated
  around $(\theta=\theta_0,\phi\leq n \phi_0)$, $ n>1$
 (see Fig. \ref{clacross}).

\vm

The  order of magnitude for the total  classical  production cross section
 can be estimated from

\begin{eqnarray}
\sigma (\pi_L) &\sim&  2\pi\int_{b_{cr}}^a P(b) bdb\nonumber\\
P(b) & =&  K V_{ph} X(b) \nonumber\\
X(b)&=&\int \vert A\vert ^2 d\Omega\nonumber\\
K&=& (Z_1Z_2\alpha)^2 (\frac{\alpha}{2\pi})^2
(\frac{m(\pi_L)}{f_{\pi_L}})^2\frac{1}{2^{16}\pi^{14}}
\nonumber\\
f_{\pi_L}&=& 7.9 \ keV \nonumber\\
 V_{ph}&\simeq& \frac{1}{6}((v_{cm}+\Delta
v_{cm})^3-v_{cm}^3)\sim  5.5 \cdot 10^{-5} \nonumber\\ \
\end{eqnarray}

\noindent
 $\vert A\vert^2$ is the obtained from leptopion production probablity
by extracting the
the coefficient $K$: $\vert A\vert^2$ has been estimated
for single velocity $v$ since the variation of $\vert A\vert^2$ with $v$ is
rather slow. $f_{\pi_L}$ has been deduced from ortopositronium decay width.
 $Z_1=Z_2=92$ (U-U collision)
 has been
assumed.
 For phase space volume factor  $V_{ph}$ it has been assumed
$v_{cm}\sim .1 $ and $\Delta v_{cm}\sim  0.1\cdot v_{cm}$.
The lower impact parameter cutoff has been assumed to be $b_{cr}=4\cdot
10^{-14} \ m$ and upper impact parameter cutoff $a$  is varied between
$10^{-12}-10^{-11}$ meters.

  \vm

The values of classical leptopion production cross section for $a=10^{-12} \ m$
and $a=10^{-11} \ m$ are $.5\cdot 10^{-5} \ \mu barn$ and $.3\cdot 10^{-3} \
\mu barn$ respectively. Classical leptopion production cross section is
by  several  orders of magnitude smaller
 than the measured $e^+e^-$ production cross section
of order $5  \ \mu b$. It turns out that in quantum model constructive
interference
at peak  for different values of impact parameter cures this disease:
mechanism
 is  analogous to quantum coherence ($\vert A_{coh}\vert^2 \propto N^2$
instead
 of $\vert A_{incoh}\vert^2\propto N$).

\subsection{Quantum  model for leptopion production}

There are good reasons for considering the quantum model.
First, the leptopion production cross section is by several orders of
 magnitude
too small in classical model. Secondly, in
 Th-Th collisions there are indications about the  presence of two velocity
peaks
 with separation $\delta \beta/\beta \sim 10^{-2}$ \cite{Cowan}
and this suggests that quantum mechanical diffraction effects might be
 in question.
These effects could come from the upper
and/or lower length
scale cutoff and from the delocalition of the wavefunction of
incoming nucleus.

\subsubsection{Formulation of the quantum model}

The formulation of the quantum model is based on very simple rule.
In  the classical model the production cross section is product of
differential cross section $d\sigma = 2\pi b db$  for  the incoming nucleus
 to scatter in a  given  solid angle element multiplied with
the differential probability $dP(b)$ to create a leptopion. In quantum model
the amplitude to create leptopion is the amplitude for
incoming nucleus to scatter with given impact parameter value multiplied
by the amplitude to create leptopion.  The product of amplitudes is taken in
x-space. For the  differential production cross section and
 production amplitude one obtains in Born approximation
the expression

\begin{eqnarray}
d\sigma&= & \vert f_B\vert^2 d\Omega \frac{d^3p}{2E_p(2\pi)^3}\nonumber\\
f_B&=&- \frac{m_R}{4\pi}\int exp(i\Delta k\cdot r) V(z,b)A(b) bdbdz d\phi
\nonumber\\ V(z,b)&=&\frac{Z_1Z_2\alpha_{em}}{r}
\end{eqnarray}

\noindent  where $\Delta k$ is the momentum exchange in Coulomb scattering
and a vector in the  scattering plane. Effectively the Coulomb potential is
replaced with
 the product
of the Coulomb potential and leptopion production amplitude $A(b)$.

\vm

The scattering amplitude can be reduced to  simpler form by using
the defining integral representation of Bessel functions

\begin{eqnarray}
f_B&=& K_0 \int F(b) J_0(\Delta kb) A(b)bdb
\nonumber\\
F(b\geq b_{cr})&=& \int dz \frac{1}{\sqrt{z^2+b^2}} =
2ln(\frac{\sqrt{a^2-b^2} + a}{b})  \nonumber\\
K_0&=& -2\pi^2 m_R Z_1Z_2\alpha_{em}\nonumber\\
\Delta k&=& 2ksin(\frac{\alpha}{2}) \ \ \ k =M_R\beta \nonumber\\
M_R&\simeq& A_Rm_p  \ \  \ A_R=\frac{A_1A_2}{A_1+A_2}
\end{eqnarray}

\noindent  where the length scale cutoffs in various integrations
 are not written explicitely.

\vm

 The presence of the impact parameter cutoffs implies
that the arguments of Bessel function is large and in a satisfactory
approximation one can use in the region of physical interest
the approximate trigonometric representation for Bessel
functions

\begin{eqnarray}
J_0(x)&\simeq & \sqrt{\frac{2}{\pi x}}cos(x-\frac{\pi}{4})
\end{eqnarray}

\noindent holding true for  large values of $x$.

\subsubsection{Calculation  of the leptopion production amplitude in the
quantum model}

The
details related to  the  calculation  of the  production amplitude
 can be found in appendix and it suffices
to describe only the general  treatment here.
The production  amplitude of the quantum model contains integrations over
the impact parameter and angle parameter $\psi$ associated with the cut.
The  integrands
appearing in the definition of the contributions $CUT_1$ and $CUT_2$
to the scattering amplitude have  simple exponential dependence on
impact parameter.  The function $F$ appearing in the definition
of the scattering amplitude is a  rather slow varying
function as compared to the Bessel function, which allows trigonometric
approximation.
This motivates the division of the  impact parameter  range into pieces so
that $F$ can approximated with its mean value inside each piece so
that integration over cutoff parameters can be performed exactly inside
 each piece.

\vm

$CUT_1$ becomes also sincular at $cos(\theta)= \beta/v_{vm}$,
 $cos(\psi)=\sin(\phi)$.
The singular contribution of the production amplitude can be extracted by
putting $cos(\psi)= sin(\phi)$ in the arguments of the exponent functions
appearing in the amplitude so that one obtains a rational function of
$cos(\psi)$ and $sin(\psi)$ integrable analytically. The
 remaining
nonsingular contribution can be integrated numerically.

\subsubsection{Dominating contribution to production cross section
and diffractive effects}

Consider now the behaviour of the
dominating  singular contribution to the production amplitude  depending on
$b$ via the exponent factor.  This  amplitude factorizes into a product

\begin{eqnarray}
f_B(sing)&=&K_0a^2 B(\Delta k) A(sing)\nonumber\\
B(\Delta k) &=&
\int F(ax) J_0( \Delta kax)exp(-\frac{sin(\phi)}{sin(\phi_0)}x)x dx
\nonumber\\
&\sim& \sqrt{\frac{2}{\pi\Delta  ka}}
\int F(ax) cos(\Delta kax-\frac{\pi}{4})
exp(-\frac{sin(\phi)}{sin(\phi_0)}x)\sqrt{x} dx
\nonumber\\
x&=& \frac{b}{a}
\end{eqnarray}

\noindent  The factor $A(sing)$ is the analytically calculable  singular
 and dominating  part
of the leptopion production amplitude
(see appendix)
 with the  exponential factor excluded. The factor $B$ is responsible for
diffractive effects. The contribution of the peak to the total
 production cross section is
of same order of magnitude as the classical production cross section.

\vm

At the peak $\phi\sim 0$ the contribution
 the exponent of
the  production amplitude is constant at this limit  one obtains product
of the Fourier transform of Coulomb potential with cutoffs with the
production amplitude. One can calculate the Fourier transform of the Coulomb
potential analytically to obtain

\begin{eqnarray}
f_B(sing)&\simeq &4\pi K_0
\frac{ (cos(\Delta ka)-cos( \Delta k b_{cr}))}{ \Delta k^2}
CUT_1\nonumber\\ \Delta k&=&2M_R\beta sin(\frac{\alpha}{2})
\end{eqnarray}

\noindent  One obtains oscillatory behaviour as a function
of the collision velocity in fixed angle scattering and
the period of oscillation depends on scattering angle and varies in wide
limits.

\vm

The relationship between scattering angle $\alpha$ and impact parameter
in Coulomb scattering translates the impact parameter cutoffs to the
scattering angle cutoffs

\begin{eqnarray}
a&=& \frac{Z_1Z_2\alpha_{em}}{M_R\beta^2} cot(\alpha (min)/2)
\nonumber\\
b_{cr}&=& \frac{Z_1Z_2\alpha_{em}}{M_R\beta^2} cot(\alpha (max)/2)
\end{eqnarray}

\noindent This gives for the argument $\Delta kb$ of the Bessel function
at lower and upper cutoffs the approximate  expressions

\begin{eqnarray}
\Delta ka&\simeq& \frac{2Z_1Z_2\alpha_{em}}{\beta} \sim \frac{124}{\beta}
\nonumber\\ \Delta kb_{cr}&\simeq & x_0  \frac{2Z_1Z_2\alpha_{em}}{\beta}
\sim  \frac{124x_0}{\beta}
\end{eqnarray}

\noindent The numerical values are for $Z_1=Z_2=92$ (U-U collision).
What is remarkable that the argument $\Delta ka$ at upper momentum
 cutoff does not depend
at all on the value of the cutoff length.
 The resulting oscillation at minimum scattering angle
is more rapid than allowed by
the width of the observed peak: $\Delta
\beta/\beta \sim  3\cdot 10^{-3}$ instead of  $\Delta
\beta/\beta \sim   10^{-2}$:  of course,    the measured value
need  not correspond to minimum scattering angle.
  The oscillation associated
 with the lower cutoff  comes from  $cos(2M_Rb_{cr}\beta sin(\alpha/2))$
and is slow  for   small scattering
angles $\alpha <1/A_R \sim 10^{-2}$.
For  $\alpha (max)$ the oscillation is rapid: $\delta \beta/\beta
\sim 10^{-3}$.

\vm

In the total production cross section integrated over all scattering angles
(or finite angular range) diffractive
effects disappear. This might explain why
the peak has not been observed in
some experiments \cite{Cowan}.

\subsubsection{Cross sections in quantum model}

 In figure \ref{qucross}   the quantity

\begin{eqnarray}
Y&=&V_{ph}sin(\alpha)\frac{d^2\sigma}{d\Omega d\cos(\alpha)dv}\nonumber\\
 V_{ph}&\simeq& \frac{1}{6}((v_{cm}+\Delta
v_{cm})^3-v_{cm}^3)\sim  5.5 \cdot 10^{-5}
\end{eqnarray}

\noindent having same order of magnitude as total production cross section
evaluated   at minimum  scattering angle $\alpha (min)$  in quantum model
 is plotted as a function of leptopion
angle  variables $(\theta, \phi)$ for U-U collision.
 The values of the  various parameters are
 $Z_1=Z_2=92$, $ b_{cr}=4\cdot 10^{-14} \ m$, $(\beta,v) =(.102 ,.106 )$,
$(\theta_0,
\phi_0)=(16.0 ,.002 ) \ degrees$. The upper cutoff has values
 $a=10^{-12} \
m$ and $a=10^{-11} \ m$.
  Differential
cross section is concentrated on small values of $\phi$ and has  a peak at
$\theta_0$.  There is however a sizable
contribution from other values of $\theta$ in the cross section as the plot
 of
differential production cross section (see Fig. \ref{clacross}) shows. In
particular, production cross section  has peak at $\theta=\pi$, whose height
increases with    $a$.

\vm

 An upper bound for the  total leptopion production cross section is  given
by  $\sigma_{tot}\leq \int Y d\Omega$.
 Actual cross section is
expected to be smaller by a numerical factor not smaller than $1/10$.
The order of magnitude estimate for the
leptopion production cross section in quantum model is by several
orders of magnitude larger than classical cross section. The reason
is the constructive interference for the contributions of various impact
parameter values to the amplitude at the peak. The upper bounds
are summarized in table 3 for various cases: the general order of mangitude
for production cross section is one $\mu barn$.

\vm

The value of $e^+e^-$
 production cross section can be estimated as follows.
 $e^+e^-$ pairs are produced from
via the creation of $\sigma_L\pi_L$ pairs from vacuum and subsequent
decay $\sigma_L$ to $e^+e^-$ pairs.
The   estimate for  (or rather for the upper bound of)
  $\pi_L\sigma_L$ production cross section
is obtained as

\begin{eqnarray}
\sigma (e^+e^-)&\simeq & X\sigma (\pi_L) \nonumber\\
X&=& \frac{V_2}{V_1}(\frac{k m_{\sigma_L}}{m^2_{\pi_L}})^2
\nonumber\\
\frac{V_2}{V_1}&=&V_{rel}=\frac{v_{12}^3}{3(2\pi)^2}
\sim 1.1 \cdot 10^{-5}\nonumber\\
\frac{k}{m_{pi_L}}&=&\frac{(m_{\sigma}^2-m_{\pi_L}^2)}{2m_{\pi_L}f_{\pi_L}}
\end{eqnarray}

\noindent Here  $V_2/V_1$
of two-particle and single particle phase space volumes.
$V_2$  is in good approximation
the product $V_1(cm)V_1(rel) $ of single particle phase space volumes
associated with cm coordinate and relative coordinate and one has
$V_2/V_1\sim V_{rel}= \frac{v_{12}^3}{3(2\pi)^2)} \simeq 1.1 \cdot 10^{-5}$
 if the maximum value of
the relative velocity is  $v_{12}\sim .1$.
 Situation is saved by the anomalously large value of
 $\sigma_L\pi_L\pi_L$ coupling constant $k$  appearing in
the  production vertex $k \sigma_L\pi_L\pi_L(class) $.

\vm

The resulting upper bound for the
cross section is given  in table 3 in 3 cases: the  actual cross section
contains a numerical factor not smaller than 10.  Production cross section is
very sensitive to the value of $f_{\pi_L}$ and Op anomaly $\Delta
\Gamma/\Gamma =5\cdot 10^{-3}$ gives upper bound $2 \ \mu b$ for
$a=10^{-11} \ m$, which is smaller than the experimental upper bound
$5 \ \mu b$. The lacking factor of order $5$ could come from several sources
(phase space volume, sensitive depenendence of $f_{\pi_L}$ on the mass of
the lightest leptopion,etc...).

\vm

    It must be emphasized that the
estimate is very rough  (the  replacement of integral over the angle $\alpha$
with rough  upper bound,
 estimate for the phase space volume, the values of
cutoff radii, the neglect of the velocity dependence of
the production cross section, the estimate for the minimum scattering angle,
...). It seems however safe to conclude that correct
value of the  production cross section can be reproduced with a  suitable
finetuning of the cutoff length scales $b_{cr}$ and $a$.
\vl

\begin{tabular}{||c |c | c|c |c|c|c||}\hline \cline{1-7}
$N$& $Op/10^{-3}$    & $\Gamma(\pi_L)/keV$ &$\sigma (\pi_L)/\mu b$
  & $\sigma (\pi_L)/\mu b$    &$\sigma (e^+e^-)/\mu b$
&$\sigma (e^+e^-)/\mu b$ \\ \hline
   &  & &$a=.01$ & $a=.1 $ & $a=.01$ & $a=.1 $
\\ \hline
1& 1   & $.51 $  & $.13$ &$1.4$  &$.03$   &$.3$  \\  \hline
3 & 1  &  $.27$ & $ .07$& $.74$  & $.007$  &$.08$\\ \hline
3 & 5  & $1.3$  & $ .34$& $ 3.7 $&  $ .2  $ & $ 2.0$    \\ \hline \hline
\end{tabular}

\vl

Table 3. The table summarizes  leptopion lifetime and the upper bounds
for  leptopion
 and  $e^+e^-$
production cross sections for lightest leptopion.  $N$ refers to the number of
leptopion states and $Op=\Delta \Gamma/\Gamma$ refers to
ortopositronium decay anomaly. The values
of upper cutoff length  $a$ are in units of $10^{-10} \ m$.

\begin{figure}[htb]
\leavevmode
\centering
\vspace*{1cm}
\label{qucross}
\caption{Dependence of the quantity $P=V_{ph}sin^2(\alpha)\frac{d\sigma}
{d\Omega
dvdcos(\alpha)}$  on  angles $\theta$ and $\phi$ in quantum model for cutoff
cutoff angle $\alpha (min)$.
 a) Total view for $a=10^{-12} \ m$. b) Singular region for $a=10^{-12} \ m$.
c) Total view for $a=10^{-11} \ m$. d) Singular region for $a=10^{-11} \ m$.
Various parameter values are given in the text. }
 \end{figure}

\clearpage
\subsubsection{Summary}

The usefulness of the modelling leptopion production
 is that the knowledge of  leptopion production rate makes
it possible to estimate also the production rates for other leptohadrons
and even for many particle states consisting of leptohadrons using some
effective action describing the strong interactions between leptohadrons.
One can consider two basic models for leptopion production. The models
contain no free parameters unless one
regards cufoff length scales as such. Classical model predicts  the singular
 production characteristics
of leptopion. Quantum  model  predicts
several velocity peaks at fixed scattering angle
 and the distance between the peaks
of the  production cross section  depends sensitively on the value
of the scattering angle.  Production cross section depends sensitively
on the value of the scattering angle for a fixed collision velocity.
 In both  models the reduction of the
leptopion production rate  above Coulomb wall could  be undestood as
a threshold effect: for the  collisions with
impact parameter  smaller than two times nuclear radius the production
amplitude becomes very small since $E\cdot B$ is more or less random
 for  these collisions in  the interaction region. The effect is visible for
fixed sufficiently large scattering angle only.
 $e^+e^-$ production cross section is of the observed order of magnitude
 provided that $e^+e^-$ pairs originate from  the creation of
$\sigma_L\pi_L$
pairs from vacuum followed by the decay  $\sigma_L\rightarrow e^+e^-$:
radial excitations of $\sigma_L$ predicted by string model explain the
appearence of several peaks and also $\pi_L$ is predicted to have lower mass
states.

\vm

The proposed models are certainly  overidealizations: in particular the
approximation that nuclear motion is free motion fails for those values of
the
impact
 parameter, which are most important in the  classical model.
To improve the models one should  calculate the Fourier transform of
 $E \cdot B$
 using the fields
of nuclei for classical orbits in Coulomb field  rather than free motion.
The second improvement is related to the more precise modelling of the
situation at length scales below $b_{cr}$, where nuclei do not behave like
point like  charges.
A peculiar feature of the model from the point of view of standard physics
is the appearence of the classical electromagnetic fields associated with the
classical orbits of the  colliding nuclei in the definition of the quantum
model. This is in spirit with Quantum TGD: Quantum TGD associates a  unique
spacetime surface
 (classical history) to  a given 3-surface (counterpart of quantum state).

\subsection{How to observe leptonic color?}

 The most obvious  argument against leptohadrons is that their production has
not been observed in hadronic collisions. The argument  is wrong.
Anomalously large production of  low energy
 $e^+e^-$ pairs \cite{anopair,Barshay} in hadronic collisions has been
 actually
observed. The most  natural source for photons and $e^+e^-$ pairs
are leptohadrons. There are two possibilities for the basic production
mechanism.\\
a) Colored leptons result directly from the decay of hadronic gluons.
It might be  possible to  exclude this
alternative by simple order of magnitude estimates.\\
b) Colored leptons result from the decay of virtual photons. This hypothesis
is in accordance with the general idea that the QCD:s associated with
different condensate levels of  p-adic topological condensate do not
communicate. More precisely, in TGD framework leptons and quarks
correspond  to different chiralities of configuration space spinors: this
implies that baryon and lepton numbers are conserved exactly and therefore
the stability of proton. In particular,  leptons and quarks
correspond to different Kac Moody representations: important difference as
compared with typical unified theory,  where leptons and quarks
share common multiplets of the unifying group. The special feature
of TGD is that there are several gluons since  it is possible to
associate to each Kac-Moody representation gluons, which are
"irreducible" in the sense that they couple only to a single Kac Moody
representation.
It is clear that if the physical gluons are "irreducible" the
world separates into different Kac Moody representations having
their own color interactions and communicating only via electroweak
and gravitational interactions. In particular, no strong interactions
between leptons and hadrons occur. Since  colored lepton corresponds to
octet ground state of Kac-Moody
representations the gluonic color coupling between ordinary lepton and
colored lepton vanishes.

\vm

If this picture is correct then  leptohadrons are produced only via
the ordinary electroweak interactions:  at higher energies via the decay of
virtual photon to  colored lepton pair and at low energies via the emission
of leptopion by photon. Consider next various manners to observe  the effects
of lepton color.\\
\noindent a) Resonance structure in  photon photon scattering and energy
near leptopion mass is a unique signature of leptopion.   \\
b) The production of leptomesons in strong classical
 electromagetic fields (of nuclei, for example) is one possibility. There are
several important constraints for the production of leptopions in this kind
of situation.\\
i) The scalar product $E \cdot B$ must be large. Faraway
from the source region this scalar product  tends to
vanish: consider only Coulomb field. \\
ii)  The region,  where $E \cdot B$
has considerable size cannot be too small as compared with leptopion
de Broglie wavelength (large when compared with the size of nuclei for
example).  If this condition doesn't hold true  the plane wave
appearing in Fourier amplitude is essentially constant spatially
and since the fields are approximately static the Fourier component
of $E \cdot B$  is expressible as a spatial divergence, which reduces to
a surface integral over a surface faraway from the source
region. Resulting amplitude is small since fields in faraway region
have essentially vanishing $E \cdot B$. \\
iii)  If fields are exactly static, then
energy conservation prohibits leptohadron production.\\
c) Also the production of
$e^+e^-_{ex}$ pairs in nuclear electromagnetic fields is possible although
the predicted cross section is small due to the presence of two-particle
phase space factor.
One signature of $e^-_{ex}$ is emission line accompanying the decay
$e^-_{ex}\rightarrow e^- +\gamma $. The collisions of
nuclei in highly ionized (perhaps astrophysical) plasmas provide a possible
source of leptobaryons. \\
d) The interaction of quantized em field with classical
   electromagnetic
 fields is one experimental arrangement to come into mind. The simplest
arrangement consisting of linearly polarized photons with
energy near leptopion mass  plus  constant classical
em field does not however work.  The direct production of $\pi_L-\gamma$
pairs in rapidly varying classical  electromagnetic field with frequency near
leptopion mass is perhaps a more realistic possibility . \\
 e)  In
the collisions  of  hadrons, virtual photon produced in collision can
decay to two colored leptons, which in turn fragment into leptohadrons. As a
result leptohadrons are produced, which in turn produce leptopions and
leptosigmas decaying to photon pairs and $e^+e^-$ pairs.  As already noticed,
anomalous    production of low energy
 $e^+e^-$ pairs   \cite{anopair} in hadronic collisions has
been observed.  \\
f) $e-\nu_e$ and $e-\bar{nu}_e$ scattering at energies below one MeV provide
a unique signature of leptopion. In $e-\bar{\nu}_e$  scattering $\pi_L$
appears as resonance.

\section{Leptohadron hypothesis and solar neutrino problem}

TGD predicts two new effects, which might have some role in the understanding
of the peculiarities related to solar neutrinos.\\
a) The existence of classical long range   $Z^0$  fields.\\
b) Leptohadrons and neutrino electron interaction mediated by leptopion and
leptosigma   exchange (also other leptomeson exchanges are in principle
possible).

\vm

These effects might provide solution to the puzzling features associated with
solar neutrinos.\\
a) Solar neutrino deficit seems to be an established fact.
\\
 b) The  values of the measured neutrino flux vary. The value
measured in Homestake (neutrino energies above $.8 \ MeV$ is
in the range $[1/4,1/3]$ \cite{Davis}.
 The value measured in Kamiokande ($E_{\nu}>7 \ MeV$) is roughly $1/2$
\cite{Hirata1} and
the values measured in Gallex \cite{GALLEX} and Sage \cite{SAGE}
 ($E_{\nu}<.42 \ MeV$)  are
are $.63$ (Gallex) and $.44$ (Sage). Within experimental errors all
measurements  except Homestake are consistent with the value $J\sim 1/2$ of
the solar neutrino flux.

\vm
 Standard model
 explanations of solar neutrino deficit
in terms of mixing of different neutrino families seem to be excluded since
they require mass difference $\vert m_{\nu_{\mu}}^2-m_{\nu_e}^2\vert$, which
is much smaller than the mass difference $\sqrt{\Delta m^2 }\in 0.5-5 \ eV $
suggested by the Los Alamos experiment \cite{mixing}.  The upper bound
for the value of the mixing angle is so small that mixing scenarios for
standard model neutrinos
are totally  excluded.
 A potential difficulty of
the models trying to solve solar neutrino
problem assuming that neutrinos have a magnetic moment
\cite{Voloshin,Fugusita} is related to
supernova physics. The magnetic moment of the neutrino implies a considerable
chirality flip rate for the left handed neutrino produced in supernova. If
right handed neutrinos are inert they escape the supernova immediately so
that the neutrino burst from the supernova becomes shorter. The data obtained
from SN1987A \cite{Hirata2} are in accordance with the absence of right
handed neutrinos. In the present case this problem is not encountered. The
reason is that the temperature inside  supernova is so high (hundreds of
MeV:s) as compared to the mass of the leptopion that the rate for the
chirality flip by leptopion exchange is negligibly small.  In TGD the
anomalous magnetic moment of neutrino is expected to be of same order of
magnitude as in standard model (the chirality flip mechanism  proposed
\cite{Heavy} based on Thomas precession
 was based on misunderstanding).

\vm

 TGD inspired solution of the  solar neutrino problem relies
on the  classical $Z^0$ magnetic  fields associated with the strong magnetic
fields of solar convective zone (in particular magnetic fields of   sunspots).
What happens that the scattering of neutrinos  in the classical $Z^0$
magnetic
 fields of the solar convective zone causes dispersion of the original radial
 neutrino flux.
 In  classical picture $Z^0$ magnetic fields
trap   left handed component of neutrino wavepacket  to circular orbit,
 which escapes after having transformed to right
handed neutrino whereas right handed component passes through without
noticing the
 presence of the $Z^0$ magnetic field.
The neutrino flux received at  Earth is reduced by a factor $(1-r)$, where
$r$  tells  the effective fraction of solar surface covered by magnetic
 structures. $r=2/3$ gives flux $1/3$ in Homestake
 The model also suggests an  anticorrelation of the
neutrino flux
 with
sunspots. The anticorrelation has been  observed in Homestake but not in
other laboratories.

\vm

  In TGD context one can imagine two possible explanations  for the
discrepancy  between different measurements:\\
a)   Leptopion exchange implies a new contribution to neutrino electron
 scattering,
which dominates over the standard contribution at sufficiently
low energies
and  at sufficiently low energies implies that the effective
solar neutrino flux measured
using neutrino-electron scattering is larger than the actual flux.  In
Kamiokande the
neutrino-electron scattering is used to detect neutrinos whereas all the
other  experimental arrangements use neutrino nucleus scattering. Therefore
leptopion exchange might explain Kamiokande-Homestake discrepancy
as suggested in \cite{Heavy} but cannot explain
the discrepancy between Homestake and other measurements based on neutrino
nucleon interaction. In fact, consistency requires that leptopion
contribution to the neutrino nucleus scattering should be   negligible at
neutrino energies $7 \ MeV \leq E\leq 14 \ MeV$ at which
 Kamiokande
measurements were performed. In fact,  for PCAC value of leptopion
coupling this is the case but if one scales the coupling by a factor of
order $3$ the situation changes: this however leads to suspicously large
cross section at energies near one MeV.
 For neutrino laboratory energies below $1$  MeV  the leptopion contribution
to the  scattering
cross section  begins to dominate (for  $E_{\nu}=.2 \ MeV$ the cross
section is predicted to be 25 times larger than standard model cross
section)  and neutrino electron scattering via leptopion exchange   provides
a unique manner  to test  leptopion idea and possibly also to observe low
energy solar neutrinos.  Even a more
dramatic effect is the resonance contribution to $\bar{\nu}-e$ scattering at
cm energy equal to  leptopion mass.
 \\
b) The classical long range $Z^0$  fields associated with Earth
might provide an
 explanation for the anomalously low neutrino
flux  measured in Homestake.  South
Dakota is situated  much nearer to the  magnetic North Pole  than the other
laboratories so that magnetic and also $Z^0$ magnetic field is expected to be
stronger there. The hypothesis explains also the anticorrelation with the
solar wind noticed in the Homestake data. Solar wind pushes the magnetic
field lines towards Earth  and makes it stronger. The effect is largest near
North pole. In fact, in van Allen belts the current created by ions trapped
around field lines and rotating around the Earth,   leads to a decrease of
the magnetic field strength near  the Equator.

\vm

To make the discussion more quantitative consider now the total cross
sections for the scattering of left and right handed neutrinos on electrons
at solar neutrino energies.
The contribution of the standard electroweak interactions to the scattering
$\nu^1_L e_2 \rightarrow \nu^3_L e_3$ \cite{Okun} is given by the expression

\begin{eqnarray}
T_{stand}&=& -\sqrt{2} G \bar{e}_2
\gamma_{\alpha}( g_LP_L +g_RP_R)e_3
\bar{\nu}_1\gamma^{\alpha}P_L\nu_4 \nonumber\\
g_L&=& 1+2sin^2(\theta_W)\ \ \ g_R= 2sin^2(\theta_W)\nonumber\\
P_L&=& \frac{1-\gamma_5}{2}\ \ P_R= \frac{1+\gamma_5}{2}\nonumber\\
\
\end{eqnarray}

\noindent The contribution of the leptopion exchange
(see Fig. \ref{leptopi}) is given the expression

\begin{eqnarray}
T_{\pi_L}&=& -\delta^2\sqrt{2} G \bar{\nu}_1
\gamma_5e_3 \frac{m_e^2}{q^2-m^2(\pi_L)}\bar{e}_2 \gamma_5\nu_4
\end{eqnarray}

\noindent where $\delta$ parametrizes the deviation of the  leptopion
coupling from PCAC value: $g= \delta g_{PCAC}$.

\vm

 If one takes seriously
the proposed model predicting two satellite trajectories for leptopion, one
must  over all
three leptopion exchanges. In this case one has several coupling
constants $g(\pi_L(i),e,\nu)$, which are assumed to be identical.
One could however argue that PCAC hypothesis forces the sum of the
couplings to be equal to the single leptopion PCAC value: this assumption
corresponds to $\delta=1/3$.
The square of the total scattering amplitude is given by

\begin{eqnarray}
T_{Kamio}^2 &=& \vert T_{stand}\vert^2 +\sum_i\vert T_{\pi_L(i)}\vert^2
+2\sum_{i} Re(T_{stand}T^{\dagger}_{\pi_L(i)})
+\sum{i\neq j}T_{\pi_L(i)}\bar{T}_{\pi_L(j)}\nonumber\\
\vert T_{stand}\vert^2&=&
2G^2(g_R^2 (p_1\cdot p_2)^2 + g_L^2(p_2\cdot p_4)^2 -g_Lg_R (p_1\cdot p_4)
m_e^2) \nonumber\\
\vert T_{\pi_L}\vert^2  &=&  G^2\delta^4 (
 (p_1\cdot p_3)^2 + (p_1\cdot p_2)^2 -p_1\cdot p_4(p_2\cdot p_3))
m_e^2X(i)^2 \nonumber\\
2Re(T_{stand}T^{\dagger}_{\pi_L})&=& 4G^2\delta^2  (
g_L p_1\cdot p_4m_e^2 -2g_R(p_1\cdot p_3)^2) X(i) \epsilon
\nonumber\\
T_{\pi_L(i)}\bar{T}_{\pi_L(j)}&=&  G^2\delta^4 (
 (p_1\cdot p_3)^2 + (p_1\cdot p_2)^2 -p_1\cdot p_4(p_2\cdot p_3))
m_e^2X(i)X(j)
\nonumber\\
X(i)&=& \frac{1}{(q^2-m_{\pi_L(i)}^2 )}\nonumber\\
\epsilon (L)&=&1 \ \ \ \epsilon (R)=0
\end{eqnarray}

\noindent     Interference term (parameter $\epsilon$ in the formula)
is present for left handed neutrinos
whereas  for righthanded neutrinos  scattering cross section is just the
sum of leptopion and standard  cross sections.

\vm

The ratio of Kamiokande and Homestake effective  neutrino fluxes
is simply the ratio of the  corresponding cross sections
 given by

\begin{eqnarray}
\frac{\Phi_{eff} (Kamio)}{\Phi (Home)}&=&\frac{\sigma_{Kamio}}
{\sigma_{stand}}
\end{eqnarray}

\noindent
The ratio is plotted in Fig. \ref{ratio1} and \ref{ratio2} for
$\delta=1/3$ (ideal PCAC) , $\delta=1$  and $\delta=2.5$ in case of left
handed neutrino.  For $\delta =2.5$ the average value of the  ratio is near
the value $1.5= (1/2)/(1/3)$ deduced from experimental fluxes for  $7<E(\nu)
< 14 \ MeV$ at which Kamiokande measurement was carried out (figure
\ref{ratio2} c).  $\delta=2.5$ alternative is probably excluded  by the
rapid growth  of the  scattering cross section below $E_{\nu}\sim 1  \ MeV$
(figure \ref{ratio1} c).   For  $\delta =1/3$ and $\delta=1$   the
deviation of the TGD prediction from standard model cross section  is
negligible  (figure \ref{ratio2} a and
b).   For $\delta=1/3$ (ideal PCAC) the value of the  effective
flux is only by 2 per cent larger than   than standard model prediction at
$E_{\nu}=.2 \ MeV$ (figures \ref{ratio1} a).

\begin{figure}[htb]
\leavevmode
\centering
\vspace*{1cm}
\label{leptopi}
\caption{Detection of neutrinos by leptopion exchange. The corresponding
diagram involving neutral leptopion can be neglected due to the small mass
of neutrino }
\end{figure}

\begin{figure}[htb]
\leavevmode
\centering
\vspace*{1cm}
\label{ratio1}
\caption{ The ratio of  predicted $\nu_e(L)-e$ scattering cross section and
standard model cross section as a function
of neutrino laboratory energy for  $\delta =1/3$,  $\delta =1$
and   $\delta =3$ (figures a), b) and c) )  in energy range
$.2-14 \ MeV$.}
 \end{figure}

\begin{figure}[htb]
\leavevmode
\centering
\vspace*{1cm}
\label{ratio2}
\caption{ The ratio of  predicted $\nu_e(L)-e$ scattering cross section and
standard model cross section as a function
of neutrino laboratory energy for  $\delta =1/3$, $\delta =1$ and
  $\delta =3$ (figures a), b) and c) )   in  neutrino
energy range $7-14 \ MeV$.}
 \end{figure}

\clearpage

\section{APPENDIX}

\subsection{ Evaluation of leptopion production amplitude}

\subsubsection{ General form of the integral}

The amplitude for leptopion production with four momentum

\begin{eqnarray}
p&=&(p_0,\bar{p})=m\gamma_1(1,vsin(\theta)cos(\phi),vsin(\theta)sin(\phi),
vcos(\theta))\nonumber\\
\gamma_1&=&1/(1-v^{2})^{1/2}
\end{eqnarray}

\noindent is essentially the Fourier component of the instanton density

\begin{eqnarray}
U(p)&=& \int e^{ip\cdot x} E \cdot B d^{4}x
\end{eqnarray}

\noindent associated with the electromagnetic field of the colliding nuclei.

\vm

Coordinates are chosen so that target nucleus is at rest at the
origin of coordinates and colliding nucleus moves along positive z
direction in $y=0$ plane with velocity $\beta$. The orbit is
approximated with straight line with impact parameter b.

\vm

Instanton density is just the scalar product of the static electric
field E of the target nucleus and magnetic field B the magnetic
field associated with the colliding nucleus
, which is
obtained by boosting the Coulomb field of static nucleus to velocity
$\beta$. The flux lines of the magnetic field rotate around the
direction of the velocity of the colliding nucleus so that instanton
density is indeed non vanishing.

\vm

The Fourier transforms of E and B for
nuclear charge $4\pi$ giving rise to Coulomb potential
$1/r$ are given by the expressions

\begin{eqnarray}
E_{i}(k) &=&N\delta (k_0)k_{i}/k^{2}
\nonumber\\B_{i}(k)&=&N\delta (\gamma (k_0-\beta k_z))k_{j}
\varepsilon_{ijz}e^{ik_x b}/((\frac{k_z}{\gamma})^2+k_T^2)
\nonumber\\
N&=&\frac{1}{(2\pi)^2}
\end{eqnarray}

\noindent  The normalization factor corresponds to momentum space integration
measure $d^4p$.
The Fourier transform of the instanton density can be expressed as a
convolution of the Fourier transforms of E and B.

\begin{eqnarray}
U(p)&=& N_1\int E(p-k)\cdot B(k)d^{4}k\nonumber\\
N_1&=& \frac{1}{(2\pi)^4}
\end{eqnarray}

\noindent In the convolution the presence of two deltafunctions makes it
possible to integrate over $k_0$ and $k_z$ and the expression for U
reduces to a two-fold integral

\begin{eqnarray}
U(p)&=&N^2N_1\beta\gamma\int dk_x dk_yexp(ikxb)(k_xp_y-k_yp_x)/AB
\nonumber\\A&=&(p_z-\frac{p_0}{\beta})^2+p_T^2+k_T^2-2k_{T}\cdot p_{T}
\nonumber\\B&=&k_T^2+(\frac{p_0}{\beta\gamma})^2
\nonumber\\p_{T}&=&(p_x,p_y)
\end{eqnarray}

To carry out the remaining integrations one can apply residy calculus.\\
a) $k_y$ integral is expressed as a sum of two pole contributions \\
b) $k_x$ integral is expressed  as a sum of two pole contributions
plus two cut contributions.

\subsubsection{$k_y$-integration}

Integration over $k_y$  can be performed by completing the integration
contour along real axis to a half circle in upper half plane
(see Fig. \ref{kyint}).

\vm

The poles of the integrand come from the two factors A and B in
denominator and are given by the expressions

\begin{eqnarray}
k_y^{1}&=& i(k_x^2+(\frac{p_0}{\beta\gamma})^2)^{1/2}
\nonumber\\
k_y^{2}&=&p_y+i((p_z-\frac{p_0}{\beta})^2+p_x^2+k_x^2-2p_xk_x)^{1/2}
\end{eqnarray}

\noindent One obtains for the amplitude an expression as a sum of two terms

\begin{eqnarray}
U&=& 2\pi i N^2N_1 \int e^{ik_x b}(U_1+U_2)dk_x
\end{eqnarray}

\noindent corresponding to two poles in upper half plane.

\vm

The explicit expression for the first term is given by

\begin{eqnarray}
U_1&=&RE_1+iIM_1\nonumber
\\RE_1&=&(k_x\frac{p_0}{\beta}y-p_xre_1/2)/(re_1^{2}+im_1^{2})
\nonumber\\
IM_1&=&(-k_xp_y re_1/2K_1^{1/2} -p_xp_yK_1^{1/2})/(re_1^{2}
+im_1^{2})\nonumber\\
re_1&=&(p_z-\frac{p_0}{\beta})^2+p_T^2-(\frac{p_0}
{\beta\gamma})^2-2p_xk_x\nonumber\\
im_1&=&-2K_1^{1/2}p_y\nonumber\\
K_1&=&k_x^2+(\frac{p_0}{\beta\gamma})^2
\end{eqnarray}

\noindent The expression for the second term is given by

\begin{eqnarray}
U_2&=&RE_2+iIM_2\nonumber\\
RE_2&=&-((k_xp_y-p_xp_y)p_y+p_xre_2/2)/(re_2^{2}+im_2^{2})\nonumber\\
IM_2&=&(-(k_xp_y-p_xp_y)re_2/2K_2^{1/2} +p_xp_y K_2^{1/2})
/(re_2^{2}+im_2^{2})\nonumber\\
re_2&=&-(p_z-\frac{p_0}{\beta})^2+(\frac{p_0}{\beta\gamma})^2+
2p_xk_x+\frac{p_0}{\beta}y-\frac{p_0}{\beta}x\nonumber\\
im_2&=&2p_yK_2^{1/2}\nonumber\\
K_2&=&(p_z-\frac{p_0}{\beta})^2+\frac{p_0}{\beta}x+k_x^2-2p_xk_x
\end{eqnarray}

\noindent A little inspection shows that the real parts
 cancel each other:$ RE_1+RE_2=0$. A further useful result is
the identitity $im_1^2+re_1^2=re_2^2+im_2^2$ and the identity
$re_2= -re_1 +2p_y^2$.

\subsubsection{ $k_x$-integration}

One cannot perform $k_x$-integration completely using residy calculus.
The reason is that the terms $IM_1$ and $IM_2$ have cuts in complex
plane. One can however reduce the integral to a sum of ple terms
plus integrals over the cuts.

\vm

The poles of $U_1$ and $U_2$ come from the denominators
and are in fact common for the two integrands. The
explicit expressions for the pole in upper halfplane, where integrand
converges exponentially are given by

\begin{eqnarray}
re_{i}^{2}&+&im_{i}^{2}=0,\  i=1,2\nonumber\\
k_x&=&(-b+i(-b^{2}+4ac)^{1/2})/2a \nonumber\\
a&=&4p_T^2\nonumber\\
b&=&-4((p_z-\frac{p_0}{\beta})^2+p_T^2-(\frac{p_0}{\beta\gamma})^2)p_x
\nonumber\\
c&=&((p_z-\frac{p_0}{\beta})^2+p_T^2-(\frac{p_0}{\beta\gamma})^2)^{2}+
4(\frac{p_0}{\beta\gamma})^2p_y^2
p_y^2
\end{eqnarray}

\noindent  A straightforward calculation using the  previous
 identities shows that the contributions of  $IM_1$ and  $IM_2$ at pole
have opposite signs and the contribution from poles vanishes identically!

\vm

The cuts associated with  $U_1$ and $U_2$ come from the square root
terms  $K_1$ and $K_2$. The condition for the appearence of the cut
is that $K_1$ ($K_2$) is real and positive. In case of $K_1$ this
condition gives

\begin{eqnarray}
k_x&=&it, \ t\in (0,\frac{p_0}{\beta\gamma})
\end{eqnarray}

In case of $K_2$ the same condition gives

\begin{eqnarray}
k_x&=&p_x+it, \ t\in(0,\frac{p_0}{\beta}-p_z)
\end{eqnarray}

\noindent Both cuts are in the direction of imaginary axis.

\vm

The integral over real axis can be completed to an integral over
semi-circle and this integral in turn can be expressed as a sum of
two terms (see Fig. \ref{kxint}).

\begin{eqnarray}
U&=&2\pi i  N^2N_1 (CUT_1+CUT_2)
\end{eqnarray}

\noindent The first term corresponds to contour, which avoids the cuts and
reduces to a sum of pole contributions. Second term corresponds to the
addition of the cut contributions.

\vm

In the following we shall give the exprsesions of various terms in
the region $\phi\in[0,\pi/2]$. Using the symmetries

\begin{eqnarray}
A(p_x,-p_y)&=&-A(p_x,p_y)\nonumber\\
A(-p_x,-p_y)&=&\bar{A}(p_x,p_y)
\end{eqnarray}
\noindent of the amplitude one can calculate the amplitude for other values
of $\phi$.

\vm

 The integration variable for cuts is the imaginary part $t$ of complexified
$k_x$. To get a more convenient form for cut integrals one can perform a
change of the integration variable

\begin{eqnarray}
cos(\psi)&=& \frac{t}{(\frac{p_0}{\beta\gamma})}
\nonumber\\
cos(\psi) &=&\frac{t}{(\frac{p_0}{\beta}-p_z)}
\nonumber\\
\psi&\in& [0,\pi/2]
\end{eqnarray}

\noindent By a painstaking calculation one verifies that the expression for
the contribution of the first cut is given by

\begin{eqnarray}
CUT_1(x)&=& \frac{1}{2}sin(\theta)sin(\phi)\int_0^{\pi/2}
exp(\frac{-cos(\psi)x} {sin(\phi_0)})A_1d\psi
\\A_1&=& \frac{(sin(\theta)cos(\phi)+iKcos(\psi))}{X_1}
\nonumber\\X_1&=& sin^2(\theta)(sin^2(\phi)-cos^2(\psi)) +K^{2}\nonumber\\
&-&2iKsin(\theta)cos(\psi)cos(\phi)
\nonumber\\K&=&\beta\gamma(1-v_{cm}cos(\theta)/\beta)
\nonumber\\v_{cm}&=&\frac{2v}{(1+v^{2})} \ \ \
sin(\phi_0)=\frac{\beta\gamma}{bm\gamma_1}\nonumber\\
x&=&\frac{b}{a}
\end{eqnarray}

\noindent The definitions of the various auxiliary variables are given in
previous formulas.

\vm

The denominator $X_1$ vanishes,  when the conditions

\begin{eqnarray}
cos(\theta)&=&\frac{\beta}{v_{cm}}
\nonumber\\sin(\phi)&=&cos(\psi)
\end{eqnarray}

\noindent hold. In forward direction the conditions express the vanishing of
the z-component of the leptopion velocity in velocity cm frame as
one can easily realize by noticing that condition reduces to the
condition $v=\beta/2$ in nonrelativistic limit.
It turns out that the contribution of first cut in fact diverges
logarithmically in the limit $\phi=0$, which corresponds to the production of
leptopion with momentum in scattering plane and with direction angle
$cos(\theta)= \beta/v_{cm}$ .

\vm

The contribution of the second cut is given by the expression

\begin{eqnarray}
CUT_2(x) &=&(usin(\theta)sin(\phi)/2)exp(iE_1x)) \int_0^{\pi/2}
exp(-E_2x))A_2d\psi \nonumber\\
A_2&=&\frac{Y}{X_2}\nonumber\\
Y&=& sin(\theta)cos(\phi) u+icos(\psi)(w/v_{cm}+(v/\beta)sin^2(\theta)(
sin^2(\phi)-
cos^2(\phi))\nonumber\\
X_2&=&sin^2(\theta)(\frac{sin^2(\phi)}{\gamma^2} -u^{2}cos^2(\psi)\nonumber\\
&+ &\beta^2 (v^{2}sin^2(\theta)-\frac{2vw}{v_{cm}})cos^2(\phi))\nonumber\\
& +&
 \frac{w^2}{v_{cm}^2}
+2iu\beta(vsin^2(\theta)cos(\phi)-\frac{wcos(\psi)}{v_{cm}})sin(\theta)
cos(\phi)\nonumber\\ E_1&=& \frac{\gamma cos(\psi)}{sin(\phi_0) }\nonumber\\
 E_2&=&\frac{\beta\gamma vsin(\theta)cos(\phi)}{sin(\phi_0)}\nonumber\\
u&=&1-\beta vcos(\theta) \ \ \ w= 1- \frac{v_{cm}}{\beta} cos(\theta)
\nonumber\\ \
\end{eqnarray}

\noindent The denominator $X_2$ has no poles and the contribution of the
second cut is therefore finite.

\begin{figure}[htb]
\leavevmode
\centering
\vspace*{1cm}
\label{kyint}
\caption{ Evaluation of $k_y$-integral using residy calculus.}
\end{figure}

\begin{figure}[htb]
\leavevmode
\centering
\vspace*{1cm}
\label{kxint}
\caption{ Evaluation of $k_x$-integral using residy calculus.}
\end{figure}

\clearpage
\subsection{ Production amplitude in quantum model}

The previous expressions for $CUT_1$ and $CUT_2$ as such give the
 production amplitude in the classical model.
In quantum model the produciton amplitude can be reduced to  simpler
 form by using
the defining integral representation of Bessel functions

\begin{eqnarray}
f_B&=&  K_0\int F(b) J_0(\Delta kb) A(b)bdb
\nonumber\\
F(b\geq b_{cr})&=& \int dz \frac{1}{\sqrt{z^2+b^2}}=
2ln(\frac{\sqrt{a^2-b^2} + a}{b})  \nonumber\\ K_0&=& -2\pi^2 m_R
Z_1Z_2\alpha_{em}\nonumber\\ \Delta k&=& 2ksin(\frac{\alpha}{2}) \ \ \ k
=M_R\beta  \end{eqnarray}

\noindent  Note that $F$ is a  rather slowly varying function of $b$.

\vm

 The presence of the impact parameter cutoffs implies
that the arguments of Bessel function is large and in a satisfactory
approximation one can use in the region of physical interest
the approximate trigonometric representation for Bessel
functions

\begin{eqnarray}
J_0(x)&\simeq & \sqrt{\frac{2}{\pi x}}cos(x-\frac{\pi}{4})
\end{eqnarray}

\noindent holding true for  large values of $x$.
For the numerical treatment it is advantageous to perform the
integration over impact parameters before the integration over
the cut  parameter $\psi$.  One can write
the following general expression for the
contribution of the first cut to the production amplitude in quantum
model

\begin{eqnarray}
B_1&= &  K_0sin(\theta)sin(\phi)\int_0^{\pi/2}
 H (C_1)  A_1 d\psi  \nonumber\\
H(C_1)&=&  \int_0^1  F(ax)  J_0(\Delta kax)exp(-C_1ax) xdx\nonumber\\
  &\simeq& \frac{\sqrt{2}}{\sqrt{\Delta k\pi}} \int_0^1  F(ax)
exp(-C_1ax)  cos(\Delta kax-\frac{\pi}{4}) \sqrt{x}dx\nonumber\\
C_1a&=& \frac{cos(\psi)}{sin(\phi_0)}
\end{eqnarray}

\noindent Here the definition of $A_1$ can be found from the defining formula
for $CUT_1$. The corresponding expression for $ CUT_2$ reads as

\begin{eqnarray}
B_2(quant)&=& K_0usin(\theta)sin(\phi)
\int_0^{\pi/2}  H(C_2)  A_2 d\psi  \nonumber\\
\nonumber\\
C_2a&=&E_1- iE_2\nonumber\\
E_1&=& \frac{\gamma cos(\psi)}{sin(\phi_0) }\nonumber\\
 E_2&=&\frac{u\beta\gamma vsin(\theta)cos(\phi)}{sin(\phi_0)}
\end{eqnarray}

\noindent  The definition of the function $H(C)$ is same as in previous
formula. The definition of $A_2$ can be found from the defining formula
of $CUT_2$.

\subsection{ Numerical evaluation of the production amplitudes}

The numerical evaluation of the production amplitude is based on the
observation that the function $ G(x)=F(ax)\sqrt{x}$ appearing in the
definition of $H(C)$, $C=C_1,C_2$,
depends varies rather slowly in the integration range as compared to
the rapidly oscillating Bessel function, which can be approximated using
trigonometric functions. This motivates the division of
the  integration range $(x_0,1)$ of  $x$
into pieces and the approximation of  $G(x)$ with its mean value  inside
 each piece so that the remaining rapidly varying
exponent functions can be integrated
 exactly inside each piece.  This gives the following approximate
expression for the function $H(C)$

\begin{eqnarray}
H(C)&=& \frac{\sqrt{2}}{\sqrt{\Delta k\pi}}
\sum_n \langle F\rangle_n   (G(Ca, \Delta k a,x(n+1)) -
G(Ca,\Delta k a,x(n)))
\nonumber\\
\langle F\rangle_n&=& F((x(n)+x(n+1))/2)\nonumber\\
G(C, u,x)&=&  \frac{1}{\vert -iu+C\vert^2 }
(Ccos(Ê ux ) +iu sin(ux))E(Cax) \nonumber\\
E(y) &= & exp(-y)
\end{eqnarray}

\noindent  The precise definition of the  mean value of the function $F$  at
range
 $n$ is to some degree  a matter of taste.

\vm

 The appearence of the singularity at $ (\theta=\theta_0,
 cos(\psi)= sin(\phi))$
  in  the  scattering amplitude  is a complication, which is avoided by
calculating analytically the contribution of the
 singular part and numerically
the remaining nonsingular part of the amplitude. The singular part of
the amplitude can be defined as the amplitude obtained by putting
$cos(\psi)=sin(\phi)$ (the pole of denominator $X_1$)
in various  exponential factors  of the amplitude so that  a rational
function of $cos(\psi)$ and $sin(\psi)$ integrable analytically
by elementary calculus results.

\vm

In  the classical model one has
the representation

\begin{eqnarray}
CUT_1(sing,x) &=& sin(\theta)sin(\phi)exp(\frac{-sin(\phi)x}{sin(\phi_0)})
\int_0^{\pi/2} A_1d\psi/2
\\A_1&=& \frac{(sin(\theta)cos(\phi)+iKcos(\psi))}{X_1}
\nonumber\\
CUT_1(reg,x)&=& CUT_1(x)-CUT_1(sing,x)\nonumber\\
x&=&b/a
\end{eqnarray}

\noindent The notations are same as in the defining formula of $CUT_1$.

\vm

In quantum   model  the corresponding replacement
is $C_1 (cos(\psi)
\rightarrow C_1(sin(\phi) \equiv D_1$. For the exponent function $E$
appearing in the approximate integration formula the decomposition
into singular and and regular parts corresponds to the following operation

\begin{eqnarray}
E&=& exp(-C_1ax) = E_{sing}+E_{reg}\nonumber\\
  E_{sing}&=&exp(-D_1ax)\nonumber\\
 E_{reg}&=& E-E_{sing}= exp(-C_1ax)-exp(-D_1ax)\nonumber\\
C_1&=&  \frac{cos(\psi)}{sin(\phi_0)}Ê\  \ \ \
D_1 = \frac{sin(\phi)}{sin(\phi_0)}\nonumber\\
\
\end{eqnarray}

\noindent  The contribution of the second cut can be estimated numerically
as such
in both cases.

\subsection{Evaluation of the singular parts of the amplitudes}

The singular parts
of the amplitudes $ CUT_1(sing)$
and
$B_1{sing}$ are rational functions of $cos(\psi)$ and the integrals over
$\psi$ can be evaluated exactly.

\vm

In the classical model the
 expression for $A_1(sing)$ appearing as integrand in the expression
of $CUT_1(sing)$ reads as

\begin{eqnarray}
A_1(sing) &=&- \frac{1}{2\sqrt{K^2+sin^2(\theta)}} (sin(\theta)cos(\phi)A_a+
iK A_b)\nonumber\\
A_a&=& I_1(\beta,\pi/2)=
\int_{0}^{\pi/2}d\psi f_1 \nonumber\\
A_b&=& I_2(\beta,\pi/2)=
 \int_{0}^{\pi/2} d\psi f_2\nonumber\\
f_1&=& \frac{1}{(cos(\psi)- c_1)(cos(\psi)- c_2)}
 \nonumber\\
 f_2&=& cos(\psi)f_1\nonumber\\
c_1&=& \frac{-iKcos(\phi) +sin(\phi)\sqrt{K^2+sin^2(\theta)}}{sin(\theta)}
\nonumber\\
c_2&=& -\bar{c}_1
 \end{eqnarray}

\noindent  Here $c_i$ are the roots of the polynomial $X_1$ appearing
in the denominator of  the integrand.

\vm

 In quantum model the  approximate
expression for the singular contribution to the production
amplitude can be written as

\begin{eqnarray}
B_1(sing) &\simeq &    k_1 \frac{sin(\theta)sin(\phi)}
{2\sqrt{K^2+sin^2(\theta)}}
\sum_n \langle F\rangle_n (I(x(n+1))-I(x(n))\nonumber\\
I(x) &=& exp(- \frac{sin(\phi)x}{sin(\phi_0)})
 (sin(\theta)cos(\phi)A_a(\Delta ka, x)+ iK A_b(\Delta ka,x))
\nonumber\\
k_1&=& 2\pi^2 m_R Z_1Z_2\alpha_{em}
\frac{\sqrt{2}}{\sqrt{\Delta k\pi}} sin(\phi_0)\nonumber\\
\
\end{eqnarray}

\noindent The expressions for the amplitudes $A_a(k,x)$ and
$A_b(k,x)$ read as

\begin{eqnarray}
A_a(k,x) &=&   cos(kx) I_3(k,0,\pi/2)  + isin(\phi_0) k sin(kx)
 I_5(k,0,\pi/2)
  \nonumber\\
A_b(k,x) &=&  cos(kx)I_4(k,0,\pi/2)
+isin(\phi_0) k sin(kx)I_3(k,0,\pi/2)
\nonumber\\
I_i(k,\alpha,\beta)&=& \int_{\alpha}^{\beta} f_i (k) d\psi\nonumber\\
f_3(k)&=&\frac{cos(\psi)}{(cos^2(\psi)+ sin^2(\phi_0)k^2)}f_1(k)\nonumber\\
f_4(k)&=& cos(\psi)f_3(k) \nonumber\\
f_5(k)&=&\frac{1}{(cos^2(\psi)+ sin^2(\phi_0)k^2)}f_1(k)\nonumber\\
\
 \end{eqnarray}

\noindent   The expressions for the  integrals
$I_i$ as functions of the endpoints $\alpha$ and $\beta$
 can be written as

\begin{eqnarray}
I_1(k, \alpha,\beta)&=&  I_0(c_1,\alpha,\beta)-
I_0(c_2,\alpha,\beta)\nonumber\\
I_2(\alpha,\beta)&=& c_1I_0(c_1,\alpha,\beta)-c_2
I_0(c_2,\alpha,\beta)\nonumber\\
I_3&=& C_{34}\sum_{i=1,2,j=3,4} \frac{1}{(c_i-c_j)}
(c_iI_0(c_i,\alpha,\beta)-c_jI_0(c_j,\alpha,\beta))\nonumber\\
I_4&=& C_{34} \sum_{i=1,2,j=3,4} \frac{1}{(c_i-c_j)}
 ( (c_i-c_j) (\beta-\alpha)
-c_i^2I_0(c_i,\alpha,\beta)+c_j^2I_0(c_j,\alpha,\beta))\nonumber\\
I_5&=& C_{34}\sum_{i=1,2,j=3,4} \frac{1}{(c_i-c_j)}
(I_0(c_i,\alpha,\beta)-I_0(c_j,\alpha,\beta))\nonumber\\
C_{34}&=&\frac{1}{c_3-c_4}=  \frac{1}{2i kasin(\phi_0) }
\end{eqnarray}

\noindent  The parameters $c_1$ and $c_2$ are the zeros of $X_1$ as function
of $cos(\psi)$ and $c_3$ and $c_4$ the zeros  of the function
$cos^2(\psi)+ k^2a^2sin^2(\phi_0)$:

\begin{eqnarray}
c_1&=& \frac{-iKcos(\phi) +sin(\phi) \sqrt{K^2+sin^2(\theta)}}{sin(\theta)}
\nonumber\\
c_2&=& \frac{-iKcos(\phi) -sin(\phi) \sqrt{K^2+sin^2(\theta)}}{sin(\theta)}
\nonumber\\
c_3&=& i kasin(\phi_0)\nonumber\\
c_4&=& - i kasin(\phi_0)\nonumber\\
\end{eqnarray}

\noindent
 The basic integral $I_0(c,\alpha,\beta)$ appearing in the formulas
 is given by

\begin{eqnarray}
I_0(c,\alpha,\beta)&=& \int_{\alpha}^{\beta}d\psi \frac{1}{(cos(\psi)-c)}
\nonumber\\
&=& \frac{1}{\sqrt{1-c^2}}(f(\alpha)-f(\beta))\nonumber\\
f(x) &=& ln(\frac{(1+ tan(x/2)t_0)}{(1- tan(x/2)t_0)}) \nonumber\\
t_0&=& \sqrt{\frac{1-c}{1+c}}
\end{eqnarray}

\noindent From the expression of $I_0$ one discovers
that scattering amplitude has   logarithmic singularity,
 when
the condition $tan(\alpha/2)=1/t_0$  or $tan(\beta/2)=1/t_0$   is satisfied
and appears,  when $c_1$ and $c_2$ are
real. This happens at the cone $K=0$ ($\theta=\theta_0$),
when the condition

\begin{eqnarray}
\sqrt{\frac{(1-sin(\phi))}{(1+sin(\phi))}}&=& tan(x/2) \nonumber\\
x&=&\alpha  \ or  \ \beta
\end{eqnarray}

\noindent holds true. The condition is satisfied for $\phi \simeq x/2$.
$x=0$ is the only interesting case and gives singularity at $\phi=0$. In the
classical case this gives logarithmic singularity in production amplitude for
all scattering angles.

\end{document}